\documentclass[manuscript]{emulateapj}
\usepackage{color}

\begin{document}

\newcommand{\revi}{} 

\newcommand{\oiii}{[O{\sc iii}] }
\newcommand{\neiii}{[Ne{\sc iii}] }
\newcommand{\oiihb}{[O{\sc ii}]/H$\beta$ }
\newcommand{\oii}{[O{\sc ii}] }
\newcommand{\oiiihb}{[O{\sc iii}]/H$\beta$ }
\newcommand{\hb}{H$\beta$ }
\newcommand{\rrr}{${\rm R}_{23}$}
\newcommand{\ooo}{${\rm O}_{32}$}
\newcommand{\kms}{km\,s$^{-1}$}
\newcommand{\egs}{erg\,s$^{-1}$}
\newcommand{\ms}{M$_{\sun}$}
\newcommand{\sy}{M$_{\sun}$~yr$^{-1}$}
\newcommand{\oiihbew}{([O{\sc ii}]/H$\beta$)$_\mathrm{EW}$}

\newcommand {\ges} {\ {\raise-.5ex\hbox{$\buildrel>\over\sim$}}\ }
\newcommand {\les} {\ {\raise-.5ex\hbox{$\buildrel<\over\sim$}}\ } 

\newcommand{\fbr}[1]{\textcolor{magenta}{[#1]}}

\title{An Observed Link between Active Galactic Nuclei and Violent Disk Instabilities\\
in High-Redshift Galaxies}

\author{Fr\'ed\'eric Bournaud}
\affil{CEA, IRFU, SAp, F-91191 Gif-sur-Yvette, France.}
    
\author{St\'ephanie Juneau}
\affil{Steward Observatory, University of Arizona, Tucson, AZ 85721, USA.\\CEA, IRFU, SAp, F-91191 Gif-sur-Yvette, France.}
    
\author{Emeric Le~Floc'h, James Mullaney, Emanuele Daddi}
\affil{CEA, IRFU, SAp, F-91191 Gif-sur-Yvette, France.}

\author{Avishai Dekel}
\affil{Racah Institute of Physics, The Hebrew University, Jerusalem 91904, Israel.}

\author{Pierre-Alain Duc, David Elbaz, Fadia Salmi}
\affil{CEA, IRFU, SAp, F-91191 Gif-sur-Yvette, France.}

\and

\author{Mark Dickinson}
\affil{National Optical Astronomy Observatory, 950 North Cherry Avenue, Tucson, AZ 85719, USA.}

\begin{abstract}
We provide evidence for a correlation between the presence of giant clumps and the occurrence of active galactic nuclei (AGN) in disk galaxies. Giant clumps of $10^{8-9}$\,M$_\sun$ arise from violent gravitational instability in gas-rich galaxies, and it has been proposed that this instability could feed supermassive black holes (BH). We use emission line diagnostics to compare a sample of 14 clumpy (unstable) disks and a sample of 13 smoother (stable) disks at redshift $z$\,$\sim$\,0.7. The majority of clumpy disks in our sample have a high probability of containing AGN. Their [O{\sc iii}]$\lambda${\sc 5007} emission line is strongly excited, inconsistent with low-metallicity star formation alone. [Ne{\sc iii}]$\lambda${\sc 3869}  excitation is also higher. Stable disks rarely have such properties. Stacking ultra sensitive {\em Chandra} observations (4\,Ms) reveals an X-ray excess in clumpy galaxies, which confirms the presence of AGN. The clumpy galaxies in our intermediate-redshift sample have properties typical of gas-rich disk galaxies rather than mergers, being in particular on the Main Sequence of star formation. This suggests that our findings apply to the physically-similar and numerous gas-rich unstable disks at $z$\,$>$\,1. Using the observed [O{\sc iii}] and X-ray luminosities, we conservatively estimate that AGN hosted by clumpy disks have typical bolometric luminosities of the order of a few $10^{43}$\,\egs, BH growth rates $\dot{m}_{\rm BH} \! \sim \! 10^{-2}$\,M$_\sun$\,yr$^{-1}$, and that these AGN are substantially obscured in X-rays. This moderate-luminosity mode could provide a large fraction of today's BH mass with a high duty cycle ($>$10\%), accretion bursts with higher luminosities being possible over shorter phases. Violent instabilities at high redshift (giant clumps) are a much more efficient driver of BH growth than the weak instabilities in nearby spirals (bars), and the evolution of disk instabilities with mass and redshift could explain the simultaneous downsizing of star formation and of BH growth. 
\end{abstract}

\section{Introduction}

The formation of Supermassive Black Holes (BH) at the center of galaxies appears to be connected to their stellar mass assembly history. Not only is today's BH mass closely related to the stellar mass \citep{magorrian, MF01}, but also the BH accretion history appears to follow the cosmic star formation (SF) history \citep{barger01}. The bulk of both SF activity and BH growth has long been attributed to mergers, especially at high redshift \citep[e.g.,][]{elbaz-cesarsky, dimatteo05, hopkins06} but there is increasing recognition that the SF history is not dominated by merger-induced starbursts, but by continuous SF in gas-rich star forming disks \citep[e.g.,][]{E07,daddi08,rodighiero}. Models indicate that massive galaxies at high redshift obtain a large fraction of their gas from rapid cold accretion, most of which comes as smooth gas streams and very small mergers, not major mergers: the high rate of continuous gas infall maintains high fractions of cold gas in the disk and high star formation rates \citep{keres06,brooks09,dekel09,bouche10}. 

Hence, galaxy mergers likely produce the strongest starbursts \citep[e.g.,][]{tacconi08} but they appear not to dominate the cosmic star formation history. Similarly, the most rapidly accreting BHs in bright QSOs are probably fed by mergers \citep{riechers08,walter09}, but mergers do not necessarily dominate the global budget of BH growth : moderate AGNs with sub-quasar luminosities at $z \approx 1-3$ mostly reside in normally star-forming galaxies rather than in starbursting mergers \citep{mullaney11}. While it is observed, at least locally, that mergers and interactions do trigger AGN compared to isolated galaxies \citep{ellison}, mergers may not be the dominant feeding process, due for instance to the scarcity of major mergers compared to the general galaxy population. Indeed, X-ray AGN host galaxies do not show morphological merger signatures more frequently than inactive galaxies, especially at high redshift and at least for the most frequent moderate-luminosity AGNs \citep{grogin05,gabor09,cisternas11}. The X-ray signal of AGNs could be attenuated by obscuration in gas-rich mergers \citep{hopkins06}, but the most recent studies with increasing X-ray sensitivity confirm the lack of any observable AGN-merger connection \citep{schawinski11,mullaney11,kocevski11}. Non-merging galaxies also appear to dominate the X-ray luminosity and BH accretion density, especially for moderate-luminosity AGN \citep{georgakakis09}. All these observations suggest that high-redshift BH growth largely proceeds independently from major galaxy mergers. 

\medskip

Could internal processes feed BHs and AGNs efficiently in high-redshift disk galaxies? Internal, ``secular'' AGN feeding in nearby spiral galaxies mostly involve bars \citep[and embedded nuclear bars, e.g.,][]{emsellem01}: this is a slow process, expected to provide only a small fraction of the BH mass \citep{hh06}. Yet, disk galaxies at high redshift are very different from nearby spirals. They are very gas-rich, strongly unstable, often dominated by giant clumps of gas and young stars -- while local disk galaxies are dominated by bars and spiral arms, with only low-mass star forming complexes inside the spiral arms. The ubiquity of clump-dominated morphologies in high-redshift star-forming galaxies has been pointed out at various wavelengths by \citet{cowie96, E04,E07, FS06, FS09, genzel08, guo11}. These giant clumps are not just low-mass patches of star formation: they have individual masses up to a few $10^{8-9}$\,\ms~each \citep{EE05,E09,FS06,FS11}. Most clumps are not incoming satellites, but form in-situ by gravitational instabilities, as indicated by their photometric properties \citep{E07,E09}, gas kinematics \citep{shapiro08,B08,vanstarkenburg08}, and estimates of Toomre parameters $Q<1$ \citep{genzel11}.

Giant clumps are actually the most striking feature tracing the global, violent\footnote{We call ``violent'' the ring, clump, and axisymmetric instabilities in high-redshift disks, in contrast to the weaker, non-axisymmetric modes (bars and spiral arms) that dominate low-redshift disks. In the violent high-redshift instability, gravitational torques are considerably stronger and the radial gas inflow is much faster (B11).} gravitational instability in high-redshift disks. This instability results from very high gas fractions of the order of 50\% as observed by \citet{daddi08, daddi10} and \citet{tacconi10}. Theoretically, the high gas fractions and instability are explained as the natural outcome of the rapid cosmic infall of gas (\citealt*{DSC09} (hereafter DSC09); \citealt*{agertz09}; \citealt*{ceverino10}), while merger-dominated phases produce stable systems \citep{BE09}. The disks become more stable and clumpy structures are replaced by spiral arms and bars at lower redshift, as the gas fraction decreases \citep{martig12, kraljic12} and the turbulent velocity dispersion of the gas also decreases (Kassin et al., in preparation).

\medskip

It has been recently proposed that disk instability in such high redshift conditions could efficiently feed BHs. The gaseous and stellar densities in the giant clumps are so high that black holes may start forming by runaway stellar collisions and become the seeds for future supermassive BHs \citep*{EBE08}. Furthermore, the observed instability and the associated turbulent energy dissipation trigger a gas inflow throughout the disk towards the nucleus, via intense gravitational torquing \citep{gammie01,DSC09,KB10,EB10}. The central inflow rate could feed a BH at a realistic growth rate over a few 10$^{8-9}$~yr, along with bulge growth (Bournaud et al. 2011, hereafter B11). This gas inflow toward a central BH can be triggered even when giant clumps are present in the outer disk, regardless of the clumps being long-lived against stellar feedback and able to migrate radially (\citealt{KD10}, B11). The cold gas infall maintaining the instability may persist even with powerful AGN feedback \citep{dimatteo11, dubois12}.

This mechanism could hardly be directly tested in $z$\,$\approx$\,2 clumpy disk galaxies using current facilities, though. The associated AGNs should most of the time have modest intrinsic X-ray luminosities, and would be strongly obscured by the interstellar medium in these high-redshift disks, which are gas-rich and geometrically thick (B11), in addition to potential obscuration by the AGN torus itself. Instead, narrow emission lines from gas surrounding the AGN could be a better tracer of moderate BH activity in gas-rich galaxies: the associated emission region is more extended spatially, and the existence of unobscured sight lines is more likely -- the use of narrow line diagnostics to probe moderate AGNs was illustrated, for instance, by \citet{wright10}. However, the application of these diagnostics to primordial galaxies at $z \sim 2$ remains uncertain \citep[e.g.,][]{shapley}, and the main emission lines are redshifted in the near-infrared where available data are limited to few systems with low signal-to-noise spectra. The SINS near-infrared survey gathered near-infrared spectra with high signal-to-noise ratios for a sample of $z \sim 2$ star-forming galaxies, but it is mostly targeted toward galaxies that are thought to be free from AGN, on purpose \citep{FS09} -- although it does contain at least one example of AGN in a rotating clumpy disk \citep[e.g.,][]{genzel06}.

\medskip

To get around these problems, we study in this paper the presence of AGNs in clumpy disk galaxies in the intermediate redshift range $0.4 < z < 1.0$. An advantage of this intermediate redshift range is that emission line diagnostics can be used with higher confidence than at $z \sim 2$, because the mass-metallicity relation evolves at a lower rate at $z<1$ than at higher redshifts \citep[e.g.,][]{savaglio}. Another advantage is that in this redshift range, stable disks more similar to nearby spirals are also common, making it possible to directly compare violently unstable, clumpy disks to smoother, more stable systems. Clumpy galaxies in this redshift range are fully representative for the global process of violent disk instability at high redshift (as further demonstrated with our own sample in Section~2). Hence, focusing on intermediate redshifts enables us to use reliable AGN diagnostics, apply them to samples of clumpy unstable disks galaxies, and compare to more stable spiral-like galaxies at similar redshifts.

In Section~2, we build a sample of {\em Clumpy galaxies} and a reference sample of {\em Stable disks} in the GOODS\footnote{The Great Observatories Origins Deep Survey}-South field. The clumpiness is measured both visually and by computer; the samples are about mass- and redshift-matched. In Section~3, we use narrow emission line diagnostics, in particular the Mass-Excitation (MEx) diagnostic introduced by \citet[][hereafter J11]{J11}, which is useable with optical spectra up to redshift one, statistically calibrated, and robustly tested against X-ray AGN selections {\em up to} $z \! \approx \! 1$ in J11. In Section~4, we perform X-ray stacking using the deepest 4\,Ms {\em Chandra} data. The inferred AGN luminosities and BH accretion rates are discussed in Section~5.

\medskip

Throughout the paper, we assume $\Omega_\mathrm{m}=0.3$, $\Omega_\Lambda=0.7$, $H_0=70$~km~s$^{-1}$. \oiii denotes [O{\sc iii}]$\lambda$5007, [Ne{\sc iii}] denotes [Ne{\sc iii}]$\lambda$3869, and \oii denotes [O{\sc ii}]$\lambda\lambda$3726,3729, the \oii flux being the total flux of the doublet. Line ratios such as [O{\sc iii}]/H$\beta$ denotes a flux ratio, except for [O{\sc ii}]/H$\beta$ which is an equivalent width ratio.


\section{Clumpy Disks and Stable Disks: Sample Selection, Classification, and Analysis}

Violent instabilities and giant clumps are most frequent above redshift one. Here we select intermediate-redshift clumpy systems at $z\,\sim\,0.7$, in order to utilize robust diagnostics that can distinguish the signatures of AGNs from SF, and compare to a control sample of more stable disks that are absent from $z \geq 1$ datasets. In studies covering a large redshift range, the morphological and photometric properties of clumps in galaxies at intermediate redshift are not found to be very different from those of $z \sim 2$ clumpy disks \citep{E07, E09}. The only difference is that violent instabilities persist at lower redshift only in moderate-mass galaxies. At redshift 2, clumpy disks are observed even in systems with stellar masses of several $10^{11}$\ms \citep[e.g.,][]{genzel06,FS09}. Below redshift one, clumpy galaxies typically have baryonic masses of the order of a few $10^{10}$ \ms (e.g., Elmegreen et al.~2009a, Puech~2010). The fact that clumpy morphologies persist down to lower redshift for lower mass galaxies is consistent with the theoretical framework in which violent disk instabilities arise in gas-rich disks fed by cold streams. The persistence of the instability primarily requires the preservation of a high gas fraction in the disk (\citealt{BE09}, DSC09). Cold accretion should persist down to lower redshift for lower-mass systems \citep{DB06}. More massive galaxies form their stars and consume their gas earlier \citep[e.g.][]{J05}, indeed their gas fractions are lower \citep{kannappan04} and are unlikely to support violent instabilities. More massive galaxies also build a stellar spheroid more rapidly, which also stabilizes the disk (DSC09, \citealt{martig09},\citealt{cacciato}). 

Hence, the fact that clumpy galaxies at intermediate redshift are mostly moderate-mass systems (a few $10^{10}$ \ms) while such morphologies are also found in more massive galaxies at $z \, \sim \, 2$, does not indicate a different origin or a different dynamical evolution. This is actually a natural property of the global process of violent disk instabilities in gas-rich galaxies fed by cold gas flows. As lower-redshift systems tend to have lower gas fractions and become more stable, an external tidal field in distant interactions may sometimes be needed to trigger the instability \citep{dimatteo08,puech}, but even so the dynamical process remains a violent instability in a gas-rich disk. We will show in the following that our sample of $z\,\sim\,0.7$ clumpy galaxies are gas-rich systems on the Main Sequence of star formation, as expected for clumpy disks at higher redshift.

\subsection{Strategy and Datasets}

\begin{table*}
\centering
\caption{Properties of the galaxies in our Clumpy and Stable disk samples.\label{table1}}
\begin{tabular}{lclcccccccc}
\hline
\hline
 &&\multicolumn{3}{c}{Object}  &&  \multicolumn{2}{c}{Mass and redshift}    &&  \multicolumn{2}{c}{Clumpiness}\\
Type &$\:\:\:\:\:\:$&ID  & $\alpha$ (J2000) & $\delta$(J2000) &$\:\:\:\:$& $z_\mathrm{spec}$ & $\log(M_*/M_{\sun})$ &$\:\:\:\:$& visual & automated \\
\hline
&&S1   & 03:32:11.38 &-27:42:06.5 &$\:\:\:\:$&0.733	&10.1	&$\:\:\:\:$& 1.4 & 1.6 \\ 
&&S2   & 03:32:17.63 &-27:48:11.8 &$\:\:\:\:$&0.735	&10.7	&$\:\:\:\:$& 1.1 & 2.6 \\ 
&&S3   & 03:32:19.68 &-27:50:23.6 &$\:\:\:\:$&0.559	&10.9	&$\:\:\:\:$& 2.1 & -- \\ 
&&S4   & 03:32:23.17 &-27:55:41.2 &$\:\:\:\:$&0.733	&10.2	&$\:\:\:\:$& 2.3 & 2.4 \\ 
&&S5   & 03:32:23.40 &-27:43:16.6 &$\:\:\:\:$&0.616	&11.1	&$\:\:\:\:$& 1.6 & 2.1 \\ 
&&S6   & 03:32:27.83 &-27:43:37.5 &$\:\:\:\:$&0.548	&9.8  	&$\:\:\:\:$& 1.2 & 1.6 \\ 
Stable disks
&&S7   & 03:32:54.51 &-27:47:03.5 &$\:\:\:\:$&0.533	&10.7	&$\:\:\:\:$& 1.2 & -- \\ 
&&S8   & 03:32:08.14 &-27:47:12.3 &$\:\:\:\:$&0.578	&9.7  	&$\:\:\:\:$& 1.6 & -- \\ 
&&S9   & 03:32:20.69 &-27:51:42.1 &$\:\:\:\:$&0.679	&10.9	&$\:\:\:\:$& 2.4 & 1.6 \\ 
&&S10  & 03:32:30.43 &-27:51:40.3 &$\:\:\:\:$&0.760	&10.4	&$\:\:\:\:$& 1.1 & 2.7 \\ 
&&S11  & 03:32:29.17 &-27:48:33.1 &$\:\:\:\:$&0.432	&10.1	&$\:\:\:\:$& 1.9 & 1.4 \\ 
&&S12  & 03:32:18.69 &-27:51:49.3 &$\:\:\:\:$&0.457	&10.2	&$\:\:\:\:$& 1.4 & 1.9 \\ 
&&S13  & 03:32:19.78 &-27:54:09.1 &$\:\:\:\:$&0.735	&10.4	&$\:\:\:\:$& 1.8 & 3.1 \vspace{.02in}\\ 
&&C1   & 03:32:28.18 &-27:40:51.6 &$\:\:\:\:$&0.426	&9.9  	&$\:\:\:\:$& 4.3 & -- \\ 
&&C2   & 03:32:17.47 &-27:48:38.4 &$\:\:\:\:$&0.737	&9.8  	&$\:\:\:\:$& 4.1 & 4.5 \\ 
&&C3   & 03:32:19.61 &-27:48:31.0 &$\:\:\:\:$&0.671	&10.3	&$\:\:\:\:$& 4.2 & -- \\ 
&&C4   & 03:32:36.68 &-27:39:54.6 &$\:\:\:\:$&0.455	&10.2	&$\:\:\:\:$& 3.5 & 3.1 \\ 
&&C5   & 03:32:27.11 &-27:49:22.0 &$\:\:\:\:$&0.559	&9.7  	&$\:\:\:\:$& 4.6 & -- \\ 
&&C6   & 03:32:33.01 &-27:48:29.4 &$\:\:\:\:$&0.664	&9.6  	&$\:\:\:\:$& 4.8 & -- \\ 
Clumpy disks
&&C7   & 03:32:34.04 &-27:50:09.7 &$\:\:\:\:$&0.703	&10.2	&$\:\:\:\:$& 5.0 & -- \\
&&C8   & 03:32:51.52 &-27:47:58.1 &$\:\:\:\:$&0.737	&11.0	&$\:\:\:\:$& 3.4 & 4.7 \\ 
&&C9   & 03:32:23.66 &-27:49:38.0 &$\:\:\:\:$&0.578	&10.7	&$\:\:\:\:$& 4.4 & 4.2 \\ 
&&C10  & 03:32:15.79 &-27:53:24.7 &$\:\:\:\:$&0.676	&10.7	&$\:\:\:\:$& 4.7 & 3.8 \\ 
&&C11  & 03:32:29.52 &-27:55:27.2 &$\:\:\:\:$&0.663	&9.7  	&$\:\:\:\:$& 4.2 & -- \\ 
&&C12  & 03:32:14.59 &-27:49:13.4 &$\:\:\:\:$&0.562	&9.6  	&$\:\:\:\:$& 4.6 & 2.8 \\ 
&&C13  & 03:32:15.35 &-27:45:07.0 &$\:\:\:\:$&0.861	&10.5	&$\:\:\:\:$& 3.8 & 2.7 \\ 
&&C14  & 03:32:25.19 &-27:51:00.0 &$\:\:\:\:$&0.841	&10.4	&$\:\:\:\:$& 4.0 & 3.8 \\ 
\hline
\end{tabular}
\end{table*}

\begin{table*}
\centering
\caption{AGN probabilities for our Clumpy and Stable disk samples, for various AGN categories, using the MEx and Blue diagnostics separately.\label{table2}}
\begin{tabular}{lclcccccccccccccccc}
\hline
\hline
&& &&\multicolumn{4}{c}{$P_\mathrm{MEx}$}  && \multicolumn{4}{c}{$P_\mathrm{Blue}$} \\
Type &$\:\:\:\:\:\:$&ID  &$\:\:\:\:$& SF & comp & LINER & Sy2 &$\:\:\:\:$& SF & comp & LINER & Sy2 \\
\hline
&&S1  &&   0.49 & 0.27 & 0.23 & 0.01 && 0.17 & 0.82 & 0.00&  0.01\\
&&S2  &&   0.07 & 0.49 & 0.44&  0.00 && 0.29 & 0.71&  0.00 & 0.00\\
&&S3  &&   0.23 & 0.74 & 0.03&  0.00 && 0.71 & 0.29 & 0.00 & 0.00\\
&&S4  &&   0.86 & 0.14 & 0.00&  0.00 && 0.80 & 0.20  &0.00 & 0.00\\
&&S5  &&   0.78 & 0.22 & 0.00&  0.00 && 0.93 & 0.07  &0.00 & 0.00\\
&&S6  &&   0.92 & 0.05 & 0.03&  0.00 && 0.44 & 0.26  &0.27  &0.02\\
Stable disks
&&S7  &&   0.15 & 0.71 & 0.14&  0.00 && 0.68 & 0.31  &0.01 & 0.00\\
&&S8  &&   0.93 & 0.02 & 0.02&  0.03 && 0.58 &  0.12  &0.02  &0.29\\
&&S9  &&   0.03 & 0.40 & 0.57&  0.01 && 0.30 &  0.34  &0.34 & 0.01\\
&&S10 &&   0.12 & 0.40 & 0.47&  0.01 && 0.50 &  0.49  &0.01 & 0.00\\
&&S11 &&   0.43 & 0.31 & 0.26&  0.00 && 0.23 &  0.31  &0.44&  0.02\\
&&S12 &&   0.47 & 0.34 & 0.19&  0.00 && 0.68 &  0.29  &0.02&  0.00\\
&&S13 &&   0.89 & 0.11 & 0.00&  0.00 && 0.91 &  0.09  &0.00  &0.00\vspace{.02in}\\
&&C1   &&   0.04 & 0.01&  0.16  &0.80 && 0.07 & 0.01 & 0.00 & 0.91\\
&&C2   &&   0.80 & 0.03&  0.06  &0.11 && 0.57 & 0.09  &0.01 & 0.32\\
&&C3   &&   0.03 & 0.10&  0.53  &0.34 && 0.43 & 0.17  &0.00 & 0.40\\
&&C4   &&   0.11 & 0.12&  0.47  &0.30 &&   ---  & ---  & ---  & ---\\
&&C5   &&   0.85 & 0.02&  0.03  &0.09 && 0.43 & 0.03 & 0.01 & 0.53\\
&&C6   &&   0.96 & 0.01&  0.01  &0.02 &&   --- &  ---  & ---  & ---\\
Clumpy disks
&&C7   &&   0.02 & 0.05&  0.48  &0.45 &&   --- &  ---  &---  & ---\\
&&C8   &&   0.00 & 0.00&  0.21  &0.79 && 0.00 & 0.00 & 0.04 & 0.95\\
&&C9   &&   0.11 & 0.69&  0.20  &0.00 && 0.68 & 0.30 & 0.02 & 0.00\\
&&C10  &&   0.00 & 0.04&  0.70  &0.26 && 0.47 & 0.14 & 0.01 & 0.38\\
&&C11  &&   0.43 & 0.04&  0.11  &0.42 && 0.13 & 0.00 & 0.00 & 0.87\\
&&C12  &&   0.40 & 0.08&  0.04  &0.48 && 0.01 & 0.00 & 0.00 & 0.99\\
&&C13  &&   0.09 & 0.44&  0.46  &0.00 && 0.32 & 0.67 & 0.00 & 0.00\\
&&C14  &&   0.00  &0.00&  0.16  &0.84 && 0.00 & 0.00 & 0.04 & 0.96\\
\hline
\end{tabular}
\end{table*}

We aimed at obtaining a sample of Clumpy unstable galaxies at $z \sim 0.4-1$, and a comparison sample of Stable disk galaxies with only weak non-axisymmetric instabilities (bars and spiral arms). Having two such categories allows us to directly compare their average SF activity, AGN probabilities, and perform X-ray stacking. Beyond this binary classification, we will also use quantitative estimates of the Clumpiness, so as to distinguish the most clumpy galaxies from more moderate cases.

The parent sample was selected in the GOODS-South field, with deep HST/ACS imaging from \citet{mauro04}. This is the most suitable field combining deep-enough imaging for substructure identification (clumps), a large coverage of moderate-mass galaxies in optical spectroscopic surveys, with large-enough statistics.

The main AGN identification tool used in this study is the MEx diagnostic. This narrow emission line diagnostic is less affected by nuclear obscuration than X-ray selections, and was indeed shown to complement X-ray AGN selection with the identification of X-ray weak AGN (J11) -- X-ray stacking (Section~5) will confirm that most of the MEx-identified AGN in our sample are too faint to be individually detected in X-rays but are detected in stacked data. As opposed to the BPT technique (Baldwin, Phillips \& Terlevich, 1981), the MEx diagnostic diagram remains useable beyond $z=0.4$ with optical spectra, and its robustness was tested up to $z \approx 1$ in J11. A key requirement for our sample selection is then the availability of a reliable \oiiihb flux ratio measurement. The Blue diagram  ([O{\sc ii}]$\lambda3727$/H$\beta$$_\mathrm{EW}$ versus [O{\sc iii}]/H$\beta$, \citealt{lamareille}) will also be used whenever the \oiihbew~ratio can also be accurately measured, and the results from both AGN diagnostics will be compared.

We thus select galaxies with optical spectra covering at least the \hb and \oiii lines. We restrict this selection to the two largest optical spectroscopic surveys in GOODS-South, namely GOODS/FORS2 (\citealt{vanzella}) and GOODS/VIMOS (\citealt{popesso,balestra}). For the latter, only spectra obtained with a medium-resolution VIMOS setup were considered because low-resolution setups would not yield accurate-enough flux measurements and in most cases cover too short wavelengths. The spectra used in this work hence all have a spectral resolution $\mathcal{R} \geq 720$. We did not include other smaller spectroscopic dataset available in GOODS-South, which were examined\footnote{This was done using the entire ESO CDFS spectroscopic compilation available at {\tt http://archive.eso.org/wdb/wdb/vo/goods\_CDFS\_master}.} but would add only 2 objects each to a sample of 27 galaxies while reducing the uniformity of the spectral coverage and offering lower signal-to-noise ratio spectra.

\subsection{Sample selection}

Among all objects covered in the GOODS/FORS2 and GOODS/VIMOS surveys, we considered only those complying with the following criteria:
\begin{enumerate}
\item availability of a robust spectroscopic redshift $z_\mathrm{spec}$ with the highest quality flags in the published databases \citep{vanzella, popesso, balestra}.
\item redshift in the $0.4\!<\!z_\mathrm{spec}\!<\!1.0$ range. The upper limit is imposed by the availability of an \oiii detection, and also ensures that ACS imaging covers optical rest-frame emission. The lower limit aims at obtaining two samples of clumpy and stable galaxies that are about redshift-matched, considering that clumpy galaxies are almost absent at $z \! < \! 0.4$.
\item spectral range covering both the \oiii and \hb emission lines without being hampered by sky lines, a detection of both lines with a signal-to-noise ratio greater than four.
\item stellar mass in the $9.6\!<\!\log(M_*/{\rm M}_\odot)\!<\!11.2$ range: this arbitrary choice gives the largest possible sample without introducing a strong mass bias between clumpy and smoother disks -- we noted that in the studied redshift range, systems below our lower-mass limit are mostly clumpy, systems above our higher-mass limit are rarely clumpy. The technique employed for stellar mass estimates is described in Appendix~\ref{sec:mass}.
\end{enumerate}

\begin{figure*}
\centering
\includegraphics[width=5in]{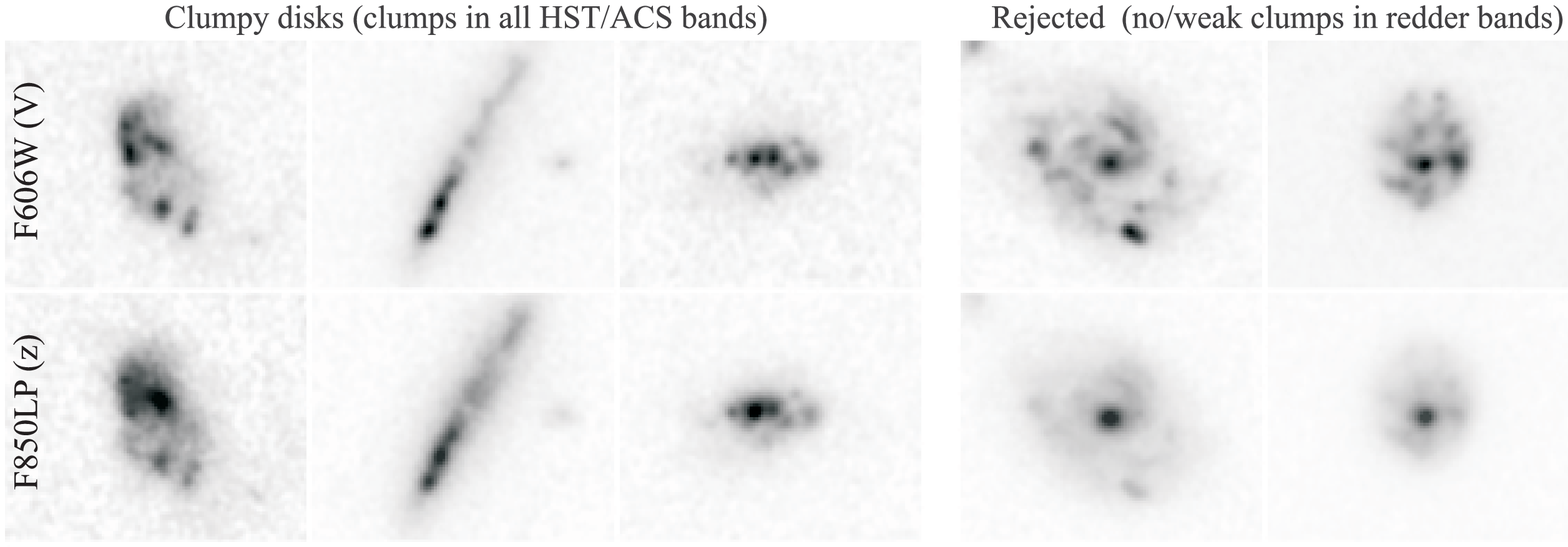}
\caption{{\revi Comparison of systems classified as ``Clumpy disks'' and rejected cases in two HST/ACS bands: the $V$ band (F606W, top) corresponding to near-UV emission for our targets, and the $z$ band (F850LP, bottom) corresponding to optical emission around the $V$ band rest-frame. Systems classified as ``Clumpy disks''  have major clumps in the optical rest-frame, even if slightly less contrasted than in the UV rest-frame. Rejected systems have weak elongated clumps that are strong in the UV rest-frame but merely visible in the optical emission. All images use the same logarithmic grey scale. \label{fig:ACSbands}}}
\end{figure*}

From this initial selection of 48 objects, we remove systems that cannot be obviously classified as Clumpy galaxies or Stable disks, namely: 

\newcounter{mynb}
\begin{enumerate}
\item compact systems with major axis smaller than $\leq0.8"$ in ACS $i$ band, as no substructure could be distinguished (6 objects).
\item interacting pairs and mergers with long tidal tails and/or double nuclei (4 objects).
\item systems with major dust lanes hampering morphological classification in optical bands (3 objects).
\item spheroid-dominated systems, selected as having an axis ratio lower than 3 in ACS $z$ and $i$ bands without having clumps or spiral arms tracing a face-on disk (4 objects).
\item systems harboring weak and/or elongated clumps, or clumps smaller than the ACS PSF, making the distinction between real clumps and short ``flocculent'' spiral arms uncertain (4 objects). {\revi These weak clumps have a high contrast only in the bluest ACS bands (ultraviolet rest-frame), and merely visible in the $z$ band (optical rest-frame), which confirms that they not similar to the massive clumps of Clumpy galaxies that are clearly visible in all ACS bands (see Fig~\ref{fig:ACSbands}).}
\setcounter{mynb}{\theenumi}
\end{enumerate}	

Hence, 29\% of the initial selection is rejected as these systems do not correspond to disk galaxies (compact objects, spheroids, major mergers), and 14\% of the systems are rejected because they could not be classified as Clumpy or Stable disks (6\%: major dust lanes, 8\%:ambiguous or unresolved substructures). Representative examples of rejected systems in each of the previous category are shown in Figure~\ref{fig:sample-rej}. This resulted in a parent sample of 27 systems with resolved morphology not corresponding to major mergers, spheroids, and for which clumps/spiral arms are large/long enough to be distinguished.
We then classified these galaxies using the classification criteria and clumpiness estimates detailed hereafter, so as to build our final ``Clumpy'' and ``Stable'' disk samples (they are displayed on Figs.~\ref{fig:sampleC} and \ref{fig:sampleS} after the classification is performed).

{\revi 
Selecting galaxies with \hb and \oiii detections can bias the sample toward AGN host galaxies. This would not hinder the comparison of Stable disks and Clumpy disks, but the absolute fractions of AGN in each sample could be increased by this selection. However, the selection is mostly biased against passive galaxies. We aim at selecting star-forming disks, which should be on the Main Sequence of star formation, and the spectroscopic surveys employed do typically detect these emission lines in such galaxies and in the mass and redshift range that we consider, without requiring AGN. Indeed, even the objects with the lowest AGN probabilities in our sample, which re mostl likely star-forming galaxies, have \oiii and \hb detected above our 4-$\sigma$ threshold (and often above 5-$\sigma$), suggesting that similar star-forming galaxies would in general be selected. }

\subsection{Morphological Classification and Clumpiness Measurements}

In order to build two samples of ``Clumpy'' and ``Stable'' disk galaxies, we performed clumpiness estimate using an eyeball classification for the whole sample, and an automated measurement for about two thirds of our galaxies that have a robust disk/bulge luminosity model available from the morphological study of GOODS-South by \citet{salmi}.

\paragraph{Visual classification}

Examining the color images from the GOODS-HST/ACS images \citep{mauro04}, each of the authors independantly attributed a grade to each galaxy in the sample, between 1 and 5, using the following description of the grades:
\begin{itemize}
\item 1.0: for robust smooth/stable disk, the stability being defined against bright clumps (spiral arms and bars, which are weaker instabilities typical for $z=0$ disks, are not taken into account),
\item 2.0: for likely smooth/stable disk (some moderate clumps but spiral-dominated),
\item 3.0: for unsure systems, mix of stable and unstable regions, or potential low bulge/disk ratio,
\item 4.0: likely clumpy unstable disk (dominated by bright clumps but significant spirals),
\item 5.0: robust clumpy/unstable disk.
\end{itemize}

In the following, we present results using the mean grade for each galaxy -- median results were also examined, and the mean grades were found to be slightly more conservative with respect to our final conclusions. We hereafter refer to the average grade for each galaxy as its {\em visual clumpiness}.

\paragraph{Automated clumpiness measurement}

A quantitative clumpiness measurement, based on bulge+disk axisymmetric luminosity model built with the {\sc Galfit} software, is available for a mass selected sample in GOODS-South from \citet{salmi}. Two thirds of our sample are covered by Salmi et al. -- objects with a too low $K$-band luminosity or insufficient {\sc Galfit} convergence are not covered. We then use these axisymmetric luminosity models, when available, to perform a clumpiness measurement which is slightly different from the one used initially by Salmi et al., in order to better distinguish clumps and spiral arms in the mass and redshift ranges that we are considering.

\begin{figure*}
\centering
\includegraphics[width=6.8in]{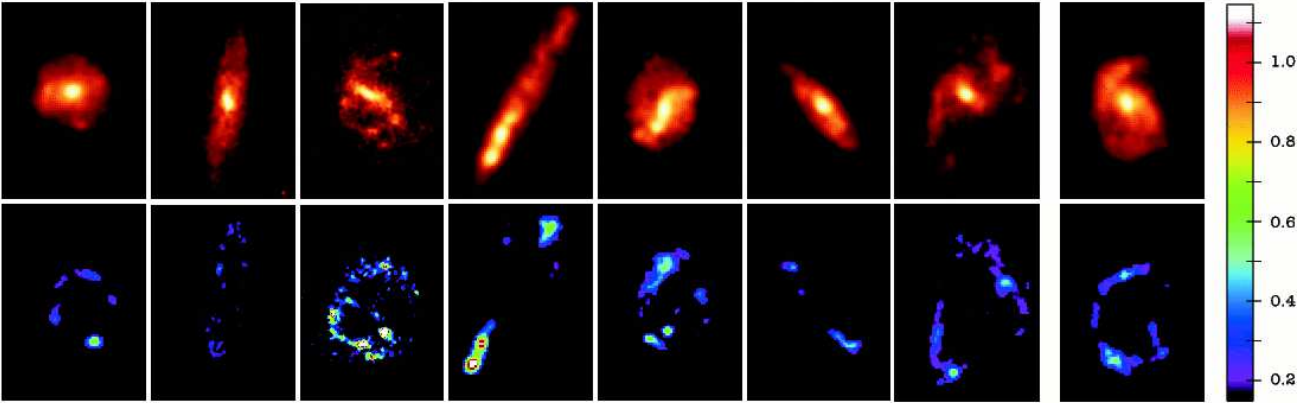}
\caption{Representative examples of ACS $z$-band images and associated relative residuals maps, using the \citet{salmi} luminosity profile models, after filtering of the relative residuals lower than 20\%, showing that spiral arms, bars and bulges are mostly removed by this 20\% threshold, while bright clumps are kept, independent of being isolated clumps or clumps along spiral arms. The last object, labeled S13 in the following sections, is the only clear case of contamination by a spiral arm that was too strong to be filtered out. The galaxies shown here are labelled C13, S11, C2, C9, C4, S10, S4 and S13, respectively, in the following sections and figures.
\label{fig:residuals}}
\end{figure*}

For each galaxy, the \citet{salmi} model was built on the $z$-band ACS image within a segmentation map. The residual map is computed as the difference between the image and model, and we divide the residual map by the model map to obtain the {\em relative residuals}. Negative residuals, which cannot correspond to bright clumps, but rather to inter-arm regions, badly subtracted bulges, or extended outer disks, are filtered out. We also filter out all relative residuals lower than 20\%, because they typically correspond to faint extended structures: spiral arms or extended outer disks, not bright clumps. Such filtering was not applied in the Salmi et al. study. Here the 20\% threshold value was found to give the best selection of ``clumps'' with respect to smoother structures, in particular spiral arms, as illustrated by the residual maps displayed in Figure~\ref{fig:residuals}. 

The {\em automated clumpiness} is defined by the sum of the residuals within the segmentation map divided by the number of pixels and multiplied by 1.8. We use the logarithm of this value, so as to obtain values more directly comparable to visual estimates. The arbitrary multiplication by 1.8 simply aims at obtaining similar scales from about 1.0 to 5.0 for both the visual and automated measurements. The correlations between clumpiness and AGN probability presented in the following Sections hold also without the use of a logarithmic scale.

\paragraph{Selection of Clumpy disk and Stable disk samples}

\begin{figure}
\centering
\includegraphics[width=2.7in]{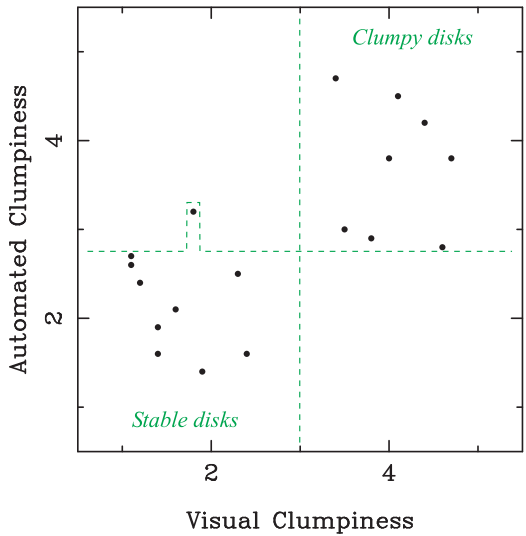}
\caption{Correlation between the visual and automated clumpiness measurements. The same samples of Clumpy and Stable disks are obtained by using the median value of either measurement -- one single object (hereafter S13) has a somewhat higher automated clumpiness value, known to be caused by a badly subtracted spiral arm, and is considered as a Stable disk.
\label{fig:visu_auto}}
\end{figure}

The automated clumpiness measurements, when available, are tightly correlated with visual estimates (Fig.~\ref{fig:visu_auto}). One single object shows a substantial difference (hereafter labeled S13), caused by a strong spiral arm that was too strong to be filtered out from the relative residual map\footnote{unless a higher filtering threshold is employed, but this would eliminate clumps in several other objects.}: the visual estimate appears more relevant for this specific case.

In order to obtain the largest possible samples of ``Clumpy'' and ``Stable'' galaxies, our main selection relies on the visual clumpiness estimates, the reliability of which was confirmed by the comparison with automated measurements. We will nevertheless present results based on the automated measurement in Section~3.6.

\begin{figure*}
\centering
\includegraphics[width=6.7in]{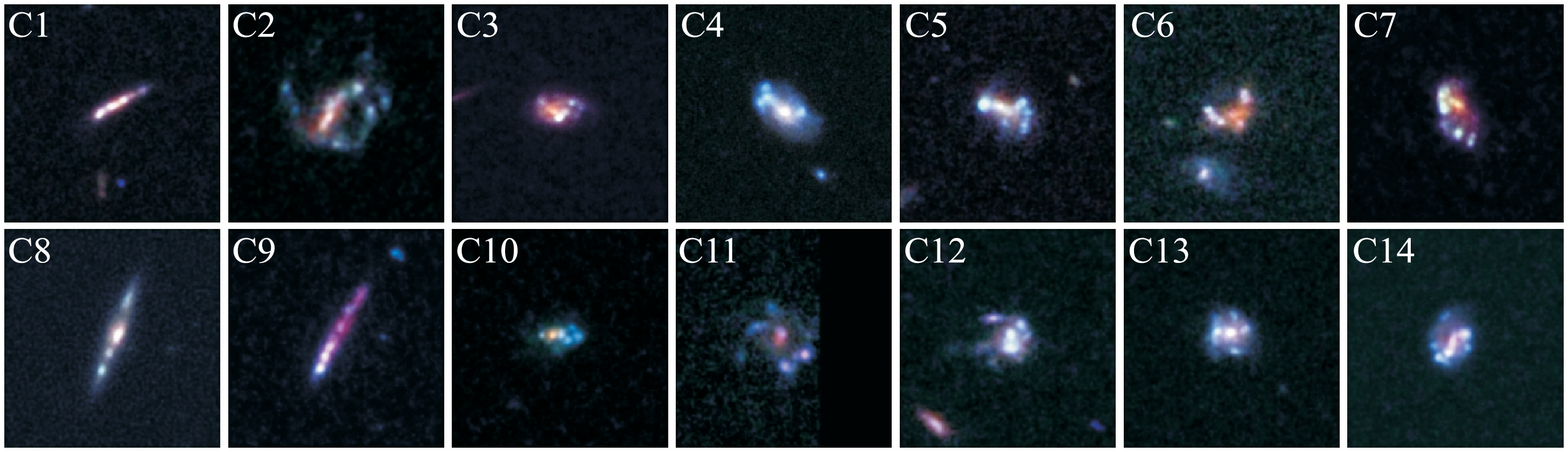}
\caption{Sample of $z \sim 0.7$ {\em Clumpy disks}. Images are 6$\times$6 arcsec, from HST/ACS $B$, $V$ and $i$ bands. Spiral arms may be present but as not as contrasted as the main clumps. \label{fig:sampleC}}
\end{figure*}

\begin{figure*}
\centering
\includegraphics[width=6.7in]{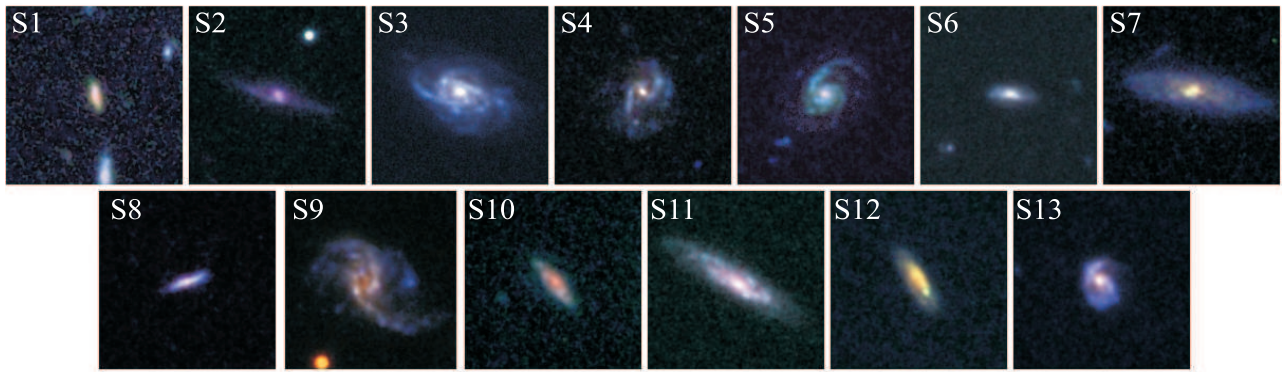}
\caption{Same as Fig.~\ref{fig:sampleC}, for our sample of {\em Stable disks}. Disk morphologies are dominated by spiral arms and bars, with weak clumps forming only in the spiral arms, indicating global stability with only local instabilities in the arms, as opposed to the ``Clumpy disk'' sample.\label{fig:sampleS}}
\end{figure*}

Here we classify the galaxies with visual clumpiness larger (respectively smaller) than 3.0 as ``Clumpy'' and ``Stable'' objects, which results in:

\begin{itemize}
\item a sample of fourteen {\em Clumpy disks} displayed in Figure~\ref{fig:sampleC} and numbered C1 to C14: their optical morphology is dominated by bright clumps, with or without spiral arms, similar to higher-redshift clumpy systems and suggesting a similar process of gravitational instability in a gas-rich disk. 
\item a sample of thirteen {\em Stable disks} displayed in Figure~\ref{fig:sampleS} and numbered from S1 to S13: their optical morphology is dominated by the usual structures found in nearby disk galaxies, mostly bars and spiral arms. Star-forming clumps can be present, in particular along the arms, but are low-luminosity ones that do not dominate compared to the arm/inter-arm contrast, as in nearby late-type spirals. The presence of clumps {\em only} along spiral arms indeed indicates that such disks are overall stable, with $Q \les 1$ instability reached only locally once the gas is compressed in the arms, as opposed to the very clumpy disks that are globally unstable with $Q \les 1$ instability not restricted to compressed spiral arm regions.
\end{itemize}

\begin{figure*}
\centering
\includegraphics[width=6.7in]{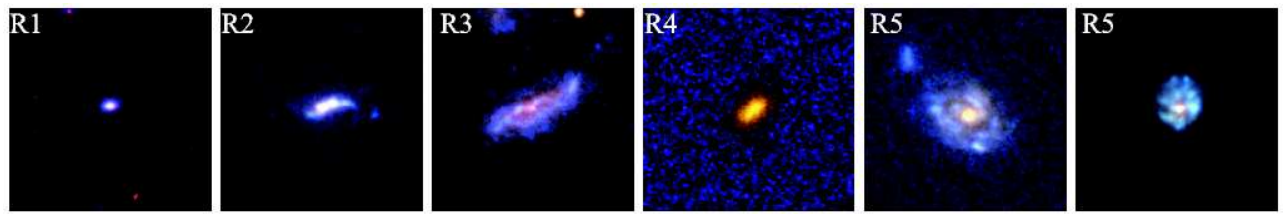}
\caption{Examples of rejected systems in our classification, because of (1) major axis smaller than 0.8 arcsec, (2) evidence for merger (double nuclei and/or tidal tails), (3) major dust absorption, (4) low axis ratio without spiral arms, suggesting spheroid-dominated nature, (5) substructures of unclear or unresolved type {\revi with weak and/or elongated clumps (two examples shown): in the first example, dust lanes seem to cut across spiral arms, resulting in apparent structures that could be either elongated clumps or short arms, in the second case the substructures are not resolved by the ACS PSF, making their nature (clump vs. short arms) unclear. In such rejected cases, the putative clumps are weak and visible only in the bluest ACS bands (i.e., ultraviolet rest-frame, see Fig.~\ref{fig:ACSbands}).} \label{fig:sample-rej}}
\end{figure*}

We note that for objects with an automated clumpiness measurement, if we separate them using the median value of these measurements, they do not change category (see Fig.~\ref{fig:visu_auto}). The only exception is galaxy S13: as already discussed, this specific case is dominated by a strong spiral arm and using its visual estimate is more relevant.

\medskip

Objects C6 and C7 were independently classified by \citet{E09}, in agreement with our present classification. These authors also noted that the last object in our Figure~\ref{fig:sample-rej} (rejected from the sample) does not show typical clumps, but rather low-mass structures that could be a combination of short flocculent spiral arms and dust absorption patches. Object S3 was classified as a regular spiral disk galaxy by \citet{neichel08}, consistent with our classification.

\medskip

Some galaxies in the Stable disk sample have bright patches or moderate clumps along their spiral arms, such as S3 or S9 (as is also the case for many nearby spirals in near-UV or blue bands). When we use a binary Clumpy/Stable classification, such objects belong to the Stable disk sample, based on their clumpiness measurements. At some point we will also compare the presence of AGN with the individual clumpiness value for each galaxy, which will fully reflect the fact that some for galaxies of the ``Stable'' sample do have some moderate clumps.

\subsection{Main properties of the Clumpy disk and Stable disk samples}

\paragraph{Mass and redshift distribution}

The Clumpy disk sample contains 14 galaxies, at a median spectroscopic redshift of 0.66 (quartiles: 0.56 and 0.74) and with a median stellar mass $\log(M_*)=10.2$ (quartiles: 9.8 and 10.6). The Stable disk sample contains 13 galaxies, at a median spectroscopic redshift of 0.62 (quartiles: 0.54 and 0.73) and with a median $\log(M_*)=10.42$ (quartiles: 10.1 and 10.8). Thus the two samples are relatively mass-matched and redshift-matched. Our samples consist of moderate-mass galaxies, although they also contain galaxies more massive than today's Milky Way. As previously detailed in Section~2, this is the expected redshift evolution of the violent instability process in galaxies fed by rapid gas inflow.

\paragraph{Star formation rates}

\begin{figure}
\centering
\includegraphics[angle=270,width=2.7in]{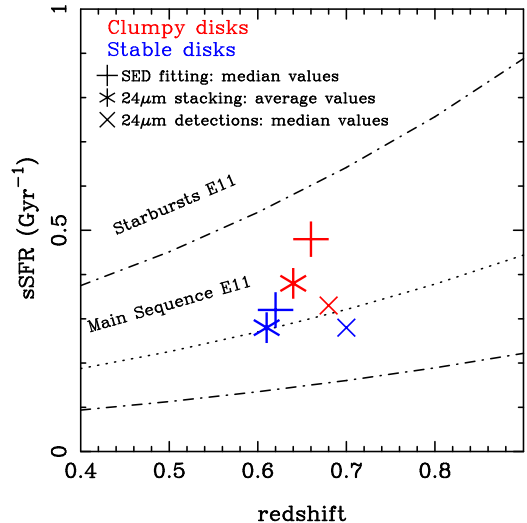}
\caption{Average or median position of our two samples in the redshift-sSFR plane, for various estimates of the star formation rate: SED fitting (median values for objects with SED fitting, see appendix), 24\,$\mu$m detections converted into SFR using \citet{chary-elbaz} (median values for about two third of each sample), and mean values from 24\,$\mu$m stacking (average values for the whole samples). The dash-dotted lines show the limits of the ``Main Sequence'' of star formation, as defined in \citet{E11}, the dotted line shows the average Main Sequence. Clumpy galaxies have higher sSFR then Stable disks, but remain within the Main Sequence, not in the ``starburst'' mode, as expected for gas-rich, gravitationally unstable disks. Note that only 24\,$\mu$m stacking cover the entire samples, while 24\,$\mu$m detections and SED fitting are available for parts of the samples, hence lying at somewhat different average/median redshifts.
\label{fig:ssfr}}
\end{figure}

Clumpy disk galaxies generally have relatively high star formation rate per unit stellar mass (sSFR, specific star formation rate), higher than stable disks and spheroid-dominated systems, but generally not as high as starbursting systems for which sSFR $>>$ 1~Gyr$^{-1}$ -- they lie within the ``Main Sequence'' of star formation defined by Elbaz et al. (2011, see also \citealt{nordon11}) that dominates the budget of star formation \citep{rodighiero}. Relatively high but not starbursting sSFRs were previously observed for $z \! \sim \! 2$ clumpy disks \citep{FS11}, and our intermediate-redshift sample has the same property, as shown on Figure~\ref{fig:ssfr}.

Individual detections in the 16 and/or 24\,$\mu$m Spitzer data from \citet{teplitz} and \citet{dickinson03}, converted into sSFR using the \citet{chary-elbaz} relations, indicate higher sSFR in Clumpy types, but only about two thirds of our targets are individually detected, with large uncertainties. To obtain a more representative estimate of the average sSFR in each sample, we stacked the 16 and 24\,$\mu$m data, using the same stacking methodology as for X-ray stacking, detailed in Section~5. None of our targets lies within the Spitzer PSF of another identified infrared source. To avoid contamination by nearby sources, we measured fluxes within the FWHM of the PSF before correcting for the full aperture.

Using these near-infrared estimates, we obtain an average sSFR in our Clumpy disk sample of $\approx 0.38$~Gyr$^{-1}$, somewhat higher than in our Stable disk sample (sSFR $\approx 0.26$~Gyr$^{-1}$). Such moderately-high sSFRs in Clumpy disks compared to Stable disks are also supported by SED fitting results (see Appendix) and by the \oii equivalent widths and H$\beta$ fluxes for Clumpy and Stable disks (after estimating and subtracting the AGN component, see Section~3.7) . We show in Figure~\ref{fig:ssfr} the position of our two samples in the redshift--sSFR plane for these various measurement techniques. This confirms that we have selected a sample of gas-rich disks with sSFRs within the upper part of the ``Main Sequence'' of star formation at $z \sim 0.7$  and not at the level of starbursting mergers \citep{E11}, and a comparison sample of more stable disks with lower sSFR, in the lower part of the ``Main Sequence''.

\medskip

The average star formation rate surface density of our Clumpy disks corresponds to a gas to total baryonic mass fraction of 35\% (using the optical major axis size for the disk diameter, and inverting the star formation - gas surface density relation for disks from Daddi et al. 2010b) . Similar estimates yield a typical gas fraction 23\% for the Stable disk sample. Applying the stability calculations presented in Bournaud \& Elmegreen (2009), we determine that violent disk instability (i.e., a Toomre parameter for gas and stars $Q \leq 1$ over most of the disk) requires a gas fraction $\ges \! 30$\%, for the average stellar mass and optical size of our sample galaxies, and assuming a bulge fraction of 20\%. The gas fractions estimated above are consistent with this instability threshold. This confirms again that we have successfully selected a sample of gas-richer, violently unstable disks (presumably representative of $z \! \sim \! 2$ clumpy galaxies), and a sample of stable disks with lower gas fractions (more representative of low-redshift spiral galaxies).


\section{Emission Line Diagnostics}

In this section we use AGN diagnostics based on stellar mass and/or emission line ratios. The methods used for stellar mass and emission line ratio measurements are detailed in the Appendix. We first study the AGN probability from the MEx diagnostic and Blue diagram, using our binary classification in Clumpy/Stable sample. We will next study the correlation between AGN probability and Clumpiness measurement for individual objects (\S 4.5).

\subsection{Optical spectra}\label{stack}

\begin{figure*}
\centering
\includegraphics[angle=270,width=11cm]{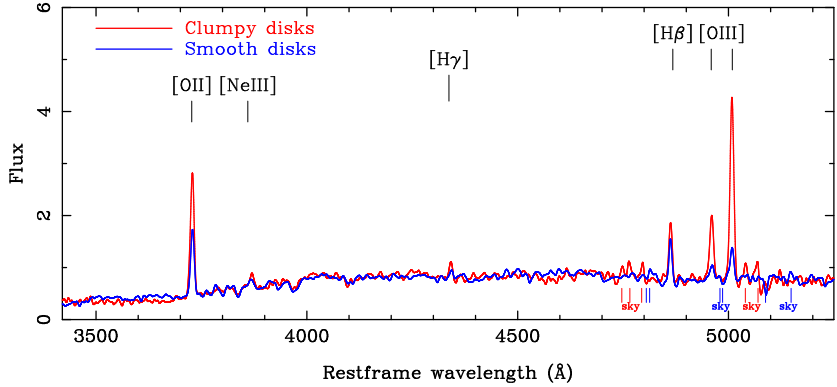}
\caption{Stacked optical spectra for our samples of ``Clumpy'' and ``Stable'' disks, normalized to the same continuum level in the 3800--4800\AA\, spectral range and to the same number of galaxies per sample, and   smoothed with a Gaussian kernel of FWHM 5\AA. Wavelengths at which significant sky lines contaminate a single spectrum, and thus potentially the stacked spectrum, are indicated: the main emission lines are left unaffected.\label{fig:spec}}
\end{figure*}

While we will use AGN diagnostics based on the individual spectra of each galaxy, we first show stacked optical spectra for the samples of Clumpy Disks and Stable Disks, so as to highlight the general properties of these two samples. We kept only galaxies for which the spectral coverage encompasses \oii as well as \oiii and \hb, and which do not have the [O{\sc iii}]$\lambda$4959 line affected by a strong sky line. This led to rejecting Clumpy systems C2, C4 and C7, as well as Stable disks S4 and S7: 78\% of the sources are included in the stacked spectra, and the excluded systems do not have extreme values of line ratios and AGN probabilities for their morphological class (see Table~2). Each individual spectrum was re-normalized to the same average continuum level in the 3800--4800\AA\,rest-frame spectral range. Stacked spectra were normalized to the same number of sources per sample. The stacked spectra, smoothed with a Gaussian kernel of FWHM 5\AA\,, are shown on Figure~\ref{fig:spec}.

The stacked spectra show that the \hb  emission, normalized to the same average continuum level, is marginally ($\approx$20\%) stronger in clumpy galaxies than in stable disks (consistent with a somewhat higher sSFR). The underlying absorption is weak in both cases, slightly higher in stable disks (as confirmed by H$\gamma$ and H$\delta$ lines). The \oii  emission is stronger by a factor 1.8 in the stacked clumpy disk spectra, while the \oiii  emission in clumpy types is a factor 5.2 stronger in clumpy disks, and the \oiiihb emission flux ratio higher by a factor 3.8. The continuum shape is quite similar in both cases. In the following sections, we use the MEx diagram and other diagnostics to show that the strong \oiii  excitation by clumpy galaxies is a likely AGN signature, and cannot be explained simply by a low gas metallicity effect.

\subsection{MEx diagnostic}

\begin{figure}
\centering
\includegraphics[width=3in]{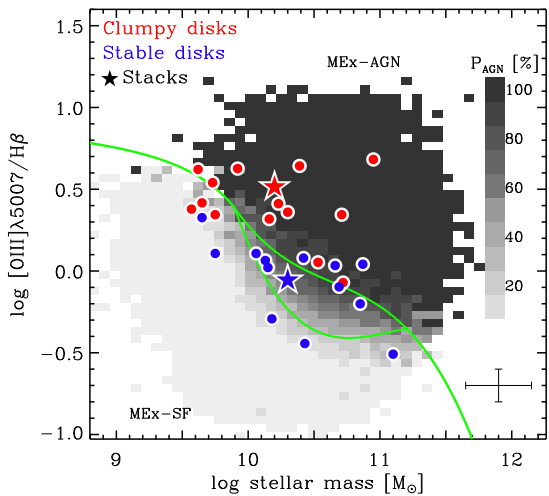}
\caption{
Clumpy and Stable disk samples on the MEx diagram. The position of the stacked spectra from Figure~\ref{fig:spec} is also shown (stars). The dividing lines separate the MEx-SF, MEx-AGN, and MEx-intermediate regions defined in Juneau et al. (2011). The gray-scaled background colors indicate the AGN probability, calibrated using the BPT diagnostic on $z\sim 0.1$ SDSS galaxies. The evolution of the mass-metallicity relation out to $z \, \approx \, 1$ in the studied mass range should not affect line ratios by more than 0.1~dex (see text Section~4.4).
\label{fig:mex}}
\end{figure}

Figure~\ref{fig:mex} shows our clumpy disk and stable disk samples on the MEx diagram (\oiiihb versus $M_*$, J11). The \oiii  excitation is clearly higher in clumpy disks, and these mostly lie above the empirical dividing line for AGN host galaxies, defined in J11. Stable disk galaxies have lower \oiii  excitations and are mostly identified as non-AGN galaxies on the MEx diagram; only a few high-mass cases lie in the MEx-intermediate or MEx-AGN regions (respectively between and above the dividing lines on Figure~\ref{fig:mex}). 

In addition to the empirical dividing lines on Fig.~\ref{fig:mex}, the MEx diagnostic was calibrated using the BPT classification of $>$$10^5$~SDSS galaxies in order to quantify the probability to observe an AGN host system as a function of \oiiihb  and $M_*$ (see J11). The SDSS calibration galaxies lie at low redshift ($z < 0.1$) and were first classified using standard BPT diagrams into the following categories: star-forming (SF), composite (comp), LINER, or Seyfert 2 (Sy2).  The likelihood of a certain spectral class is defined as the relative fraction of SDSS galaxies of that particular class within the 1$\sigma$ uncertainties on the MEx diagram. The four spectral classes listed above are mutually exclusive so their sum is, by definition, equal to unity ($P$(SF)+$P$(comp)+$P$(LINER)+$P$(Sy2)=1).  Individual AGN probabilities on the MEx diagnostic are indicated in Table~\ref{table1}. As discussed later in Section~\ref{sec-z}, the probabilities should not be strongly affected by the metallicity evolution between $z < 0.1$ (the redshift range calibration sample from J11) and $z \sim 0.7$ (the typical redshift of our samples), especially in the mass range studied here.

Stable disks have a median $P_\mathrm{MEx}(SF)$ of 47\% and, except for 2-3 AGN candidates, their AGN probabilities are low. In particular, the probability of hosting a Seyfert~2, $P_\mathrm{MEx}(Sy2)$ is always below 3\%, showing that systems that may not be pure star-forming galaxies show only weak signs of BH activity, with either composite-like properties or LINER-type excitation.

The MEx probabilities for Clumpy disks are quite different. Except for three systems where $P_\mathrm{MEx}(AGN)$ is low, the probability for the measured \oiii excitation to result purely from star formation is always below 50\%, and often close to 0. Half of the clumpy galaxies have a $P_\mathrm{MEx}(Sy2)$ comparable to or higher than the probability of composite properties or LINER-type excitation.

\subsection{Blue diagram}

\begin{figure}
\centering
\includegraphics[width=3in]{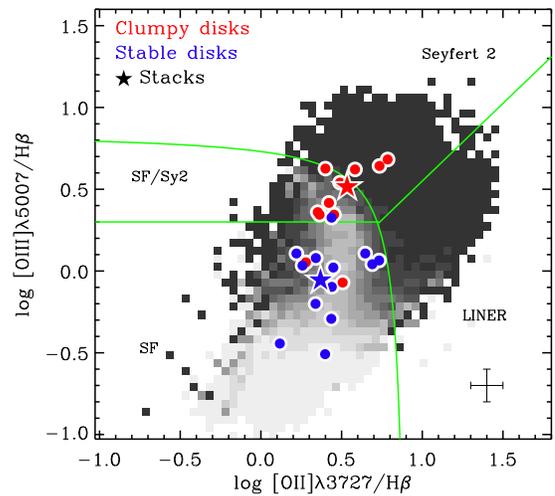}
\caption{Blue diagram with the dividing lines defined in \citep{lamareille} four our galaxy samples. The background probabilities are defined and coded as on Figure~\ref{fig:mex}. The [O{\sc ii}]/H$\beta$ ratio is here an equivalent width ratio, while [O{\sc ii}]/H$\beta$ is a flux ratio.
\label{fig:blue}}
\end{figure}

We use the Blue diagram \citep{lamareille} to perform a second classification of our samples. In some cases, the Blue diagram may be less efficient than the MEx diagnostic in separating star-forming and AGN host galaxies because of an overlap region gathering galaxies identified as star-forming and as Sy2 on the BPT diagnostic. On the other hand, the Blue diagram may more clearly separate LINER excitation (with high \oii  excitation) from Sy2 activity (with the highest excitation in \oiii). 

Our two samples are shown on the Blue diagram on Figure~\ref{fig:blue}, and Table~1 gives the associated probabilities (calibrated on SDSS galaxies, as already explained for the MEx diagnostic). This confirms that a large fraction of clumpy disks likely host an AGN, unlike the control sample of stable disks. Furthermore, the figure shows that the driving difference is the \oiii  excitation, and that clumpy disks are characterized by Sy2 or SF+Sy2 activity, rather than LINER excitation.  This feature is also apparent on the individual classification probabilities listed in Table~1, where the values of $P_{Blue}$(LINER) are all very small.

\begin{figure}
\centering
\includegraphics[width=3.4in]{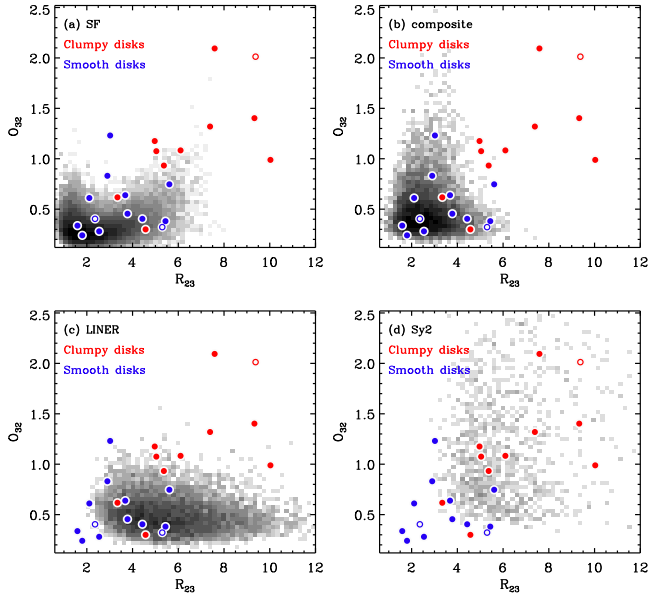}
\caption{ (\rrr ;\ooo) diagram (see text for definitions), showing the position of our two samples compared to the distribution of SDSS galaxies in four spectral classes: star-forming, composite, LINER and Seyfert~2, respectively, based on the BPT diagnostic applied to SDSS galaxies as in J11. We did not apply any dust correction to our samples and to SDSS galaxies, as we estimate the correction to O$_{32}$ to be lower than 0.1~dex.
 \label{fig:032R23}}
\end{figure}

\begin{figure*}
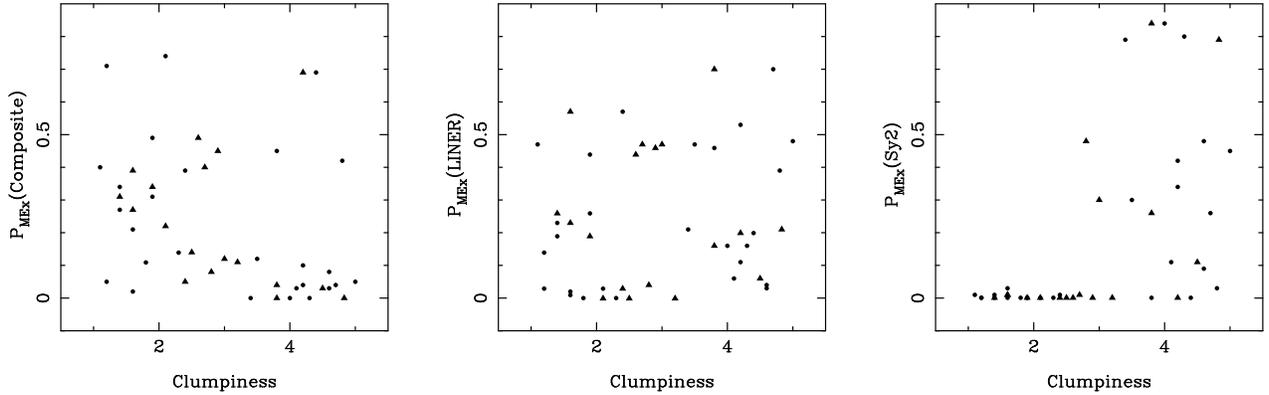

\centering
\includegraphics[angle=270,width=2in]{Pcomp_Cvis.eps}\hspace{.25in}
\includegraphics[angle=270,width=2in]{Pli_Cvis.eps}\hspace{.25in}
\includegraphics[angle=270,width=2in]{Psy2_Cvis.eps}
\caption{AGN probability versus Clumpiness value for Composite types (left), LINER types (middle), and Seyfert~2 types (right), using visual clumpiness values (circles) and automated clumpiness measurements (triangles). AGN probabilities in this figure are based on the MEx diagnostic. We indicate the corresponding linear correlation coefficients, and those based on the Blue diagram, in Table~\ref{corels}.}
\label{Pagn_vs_Clumpi}
\end{figure*}

\subsection{A metallicity effect?}\label{sec-z}

Our Clumpy galaxy sample shows, for most sources, the expected properties of AGN host systems (together with with star formation) on both the MEx and the Blue diagnostics. However, an alternative explanation for high  \oiiihb flux could be a low gas phase metallicity \citep[e.g.,][]{mcgaugh91}. Theoretically, the disk instability associated to clumpy morphologies is interpreted as being linked to the infall of fresh, low-metallicity gas (see introduction, DSC09 and Bournaud \& Elmegreen 2009), so these systems could a priori be metal-defficient. Nevertheless, simulations indicate that the gas is rapidly enriched in the dense star-forming clumps. Analysis of cosmological simulations from \citet{ceverino10} generally reveals solar metallicity in the clumps and marginally lower metallicities only in the inter-clump gas, typically one-third to half solar. Also, observations of clumpy galaxies at $z \sim 2$ do not find them to be metal-deficient \citep[e.g., ][]{genzel08, B08}. 

\medskip


As for our own sample of galaxies, the MEx diagnostic (Table~1) indicates that the probability to have the strong \oiii excitation caused by a low metallicity effect is, in general, quite lower than the AGN probability. These probabilities are calibrated on SDSS galaxies at $z < 0.1$ (J11), however the redshift evolution of the mass-metallicity relation up to $z\sim 0.7$ is not sufficient to explain the high \oiiihb ratio that we observe in Clumpy disks. For the mass range that we consider here, the typical variation of $12+\log(O/H)$ from $z<0.1$ to $z \sim 0.7$ is at most 0.1-0.2~dex, as shown by the Gemini Deep Deep Survey at $0.4<z<1.0$ by \citet{savaglio}. These authors found a slow evolution of the mass-metallicity relation with redshift at $z<1$, except for lower-mass galaxies (see also \citealt{cresci11} on the limited evolution of the mass-metallicity below redshift one). The corresponding variation of ${\rm R}_{23} \, = \, ($[O{\sc iii}]$\lambda\lambda 4959,5007 + $[O{\sc ii}]$\lambda\lambda 3726,3729 \,/ \,{\rm H}_{\beta}$ is not larger than 0.1~dex (using for instance the \rrr -metallicity relations from \citet{mcgaugh91} at fixed ionization parameter). This possible redshift evolution is insufficient to strongly affect the position of our samples on the MEx diagram, nor the computed AGN probablities, and is actually well within the assumed uncertainties.
The comparison to the Stable disk sample at similar redshift confirms that the high \oiiihb  ratios do not result from redshift evolution in the mass-metallicity relation in the studied mass range, as such an effect would affect both samples.

Furthermore, Clumpy galaxies are different from purely star-forming galaxies of any metallicity in the SDSS. This is shown by the (\rrr ;\ooo) diagrams (Fig.~\ref{fig:032R23}), where
\rrr=([O{\sc iii}]$\lambda\lambda 4959,5007$+\oii$\lambda\lambda 3726,3729)/{\rm H}_{\beta}$, 
and \ooo=[O{\sc iii}]/[O{\sc ii}]. The position of our Clumpy and Stable disk samples is compared to SDSS galaxies identified as star-forming, composite, LINER and Seyfert~2 on the BPT  diagnostic (following the analysis performed in J11). The high \rrr \, values for Clumpy galaxies come along high \ooo \, ratios. Note that high values of \rrr \, do not trace low gas-phase metallicities in case of AGN excitation. Most Clumpy disks occupy the region of the parameter space best represented by Seyfert 2's, and do not have the properties expected for star-forming galaxies of any metallicity, or for LINER excitation. Here again, the redshift evolution of the mass-metallicity relation in the studied mass range cannot explain this effect. Purely star-forming galaxies in our mass and redshift range lie almost exclusively in the \rrr$<$5 and \ooo$<$1 area of the (\rrr ;\ooo) diagram, whatever their metallicity (as also shown by \citealt{savaglio}). The majority of Clumpy disks have higher \rrr~and \ooo~ratios, and are hence different from purely star-forming galaxies of any metallicity.

\subsection{[Ne{\small \it III}]$\lambda 3869$ excitation}

\begin{figure}
\centering
\includegraphics[width=2.7in]{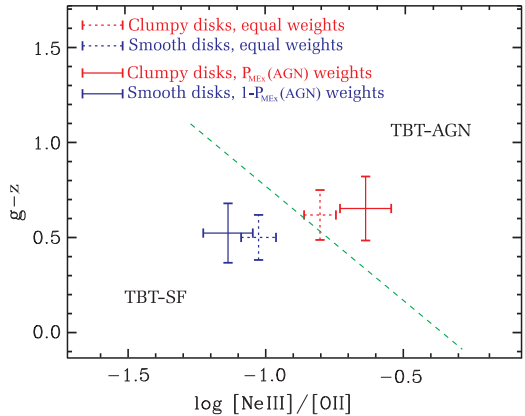}
\caption{{\revi TBT diagram for Clumpy and Stable disk samples (mean stacking) and for similar sub-samples weighted by the MEx-based probability to host AGN (for Clumpy disks) or to be purely star-forming (for Stable disks).}
 \label{fig:TBT}}
\end{figure}

{\revi

A few other AGN diagnostics based on optical lines have been proposed in the literature \citep[e.g.][]{CEx-ref}. A particularly useful one to complement the MEx diagnostic, given the spectral coverage available for our sample, is the recently proposed TBT diagnostic \citep{trouille11} that compares the \neiii$\lambda 3869$ /\oii excitation to the $(g-z)$ rest-frame color. This diagnostic typically requires to probe \neiii/\oii ratio of the order of 0.1 or greater: the spectral data available for our sample are not sensitive enough for this in individual targets, but the \neiii line is robustly detected in stacked spectral data.

The TBT diagram for our Stable Disk and Clumpy Disk samples is shown on Figure~\ref{fig:TBT}: we find that Stable Disks are dominated by purely star-forming systems, while a larger fraction of Clumpy Disks (but not necessarily all of them) should contain AGN, resulting in average properties on the AGN side of the AGN/SF limit defined by \citet{trouille11}. It is expected from the MEx diagnostic that not all Clumpy disks have optical line excitation dominated by AGN, and hence that their average properties are intermediate between purely star-forming galaxies and AGN-dominated ones, as found here on the TBT diagram. 

To further confirm the presence of AGN in many Clumpy galaxies, we have studied the stacked data for the Clumpy Disk sample with statistical weights proportional to $P_\mathrm{MEx}(AGN)$, the MEx-based probability of containing AGN, for each object. The idea is equivalent to selecting only the objects most likely to contain AGN among the Clumpy sample, improved by the use of statistical weights. The results are shown on the TBT diagram (Fig.~\ref{fig:TBT}), and the weighted data are clearly in the TBT-AGN region. Hence, while the whole sample of Clumpy Disks has intermediate properties suggesting a mix of star-forming and active galaxies, the Clumpy disks identified as likely AGN hosts by the MEx diagram are confirmed to be dominated by AGN with the TBT diagram. Conversely, the stacked data for the Stable Disk sample weighted by $P_\mathrm{MEx}(SF)=1-P_\mathrm{MEx}(AGN)$, i.e. a weighted selection of Stable Disks supposed to be likely purely star-forming according to the MEx, lie clearly in the TBT-SF region on Figure~\ref{fig:TBT}.
}

\subsection{Quantitative correlation between AGN probability and Clumpiness}

We now consider the quantitative correlation between AGN probability and Clumpiness for all individual objects, independent of the classification in two sub-samples of Clumpy and Stable disks used previously. 

We compare on Figure~\ref{Pagn_vs_Clumpi} the AGN probabilities of our objects to their visual and automated Clumpiness estimates. This is done for the three considered AGN types separately (Composite, LINER and Seyfert~2). This experiment was also performed for combinations of AGN types (such as LINER+Seyfert~2, etc), considering either the MEx diagnostic or the Blue diagram, and either the visual or the automated clumpiness estimates. In each case we determined the linear correlation coefficients $a$, $b$ and $r$ for: 
\begin{equation}
P(\mathrm{AGN}) = a \times C + b \, \pm r
\end{equation}
where $P(\mathrm{AGN})$ is the AGN probability for the considered AGN type, $C$ is the clumpiness value, and $r$ is the Pearson correlation coefficient. Results are given in Table~\ref{corels}.

\begin{table*}
\centering
\caption{Linear correlation coefficients between visual/automated clumpiness values and AGN probability, for various combinations of AGN types.}
\label{corels}
\begin{tabular}{rcccccccc}
\hline
\hline
AGN type(s)  &  $\;\;\;\;$ & \multicolumn{3}{c}{MEx} &$\;\;$& \multicolumn{3}{c}{Blue} \\
                      &          &  $a\;\:\:$ & $b\;\:\:$ & $r\;\:\:$             && $a\;\:\:$ & $b\;\:\:$ &  $r\;\:\:$                        \\
\hline
{\em Using visual clumpiness:} &&&&&&&&\\
Composite + LINER + Seyfert~2 &&  5.64  & 45.3 & 0.05   &&  8.31  & 33.6  & 0.16  \\
Composite + Seyfert~2                &&  3.46  & 29.6 & 0.04   &&  10.4  & 22.6  &  0.37 \\
LINER + Seyfert~2                      &&   13.0 & 1.90 & 0.26  &&  16.8  & -14.4  & 0.39   \\
LINER                                          &&   2.16 &  15.8 & 0.02  &&  -2.07  & 11.0  &  0.06 \\
Seyfert~2                                     &&   10.81 & -13.9 & 0.31 &&  18.9  & -25.4  & 0.47  
\vspace{.1cm}\\
{\em Using automated clumpiness:} &&&&&&&&\\
Composite + LINER + Seyfert~2 &&  8.21 & 54.9 & 0.55 && 2.19 & 46.8 & 0.07 \\
Composite + Seyfert~2                &&  9.35 & 11.8 & 0.17 && 8.30 & 26.3 & 0.28\\ 
LINER + Seyfert~2                      &&   14.49 & -0.9 & 0.18 && 14.20 & -11.9 & 0.39 \\ 
LINER                                          &&  -1.13 & 28.2 & 0.01  &&  -6.11 & 24.5 & 0.45 \\
Seyfert~2                                     &&   15.63 & -29.07 & 0.57 && 15.49 & -29.4 & 0.57 \\
\hline
\end{tabular}
\end{table*}

The global correlation between AGN probability and clumpiness is dominated by Seyfert~2-type activity: the Seyfert~2 probability vs. clumpiness correlation is the tightest one and drives the global AGN probability vs. clumpiness correlation, much more than LINER and composite types. This result is independent of the diagnostic used (MEx or Blue diagram) and preferred clumpiness measurement (visual or automated). The most clumpy objects have spectral properties that are weakly more typical of Composite galaxies compared to the less clumpy ones, and not more typical (or even less typical) of LINER galaxies, but they clearly have spectral properties that are more typical of Seyfert~2-like activity.

These results confirm the previous conclusions, independent of the chosen clumpiness threshold used to separate the {\em Clumpy} and {\em Stable } sub-samples, and of the reliability of visual clumpiness estimates. Furthermore, the fact that the strongest correlation is obtained when Seyfert~2-type activity alone is considered, and the absence of a significant correlation between LINER activity and clumpiness, confirms that the observed activity in clumpy galaxies corresponds to BH accretion, rather than shock-induced excitation or low-metallicity star formation that {\em might} have resulted in LINER-like spectral properties -- which was unlikely, as supported by the previous discussion on the evolution of the mass-metallicity relation.

It appears on Figure~\ref{Pagn_vs_Clumpi} that low-clumpiness galaxies have Composite-like properties rather than Seyfert~2-like properties, suggesting that if they contain AGN, these are lower-luminosity ones that do not dominate the [O{\sc iii}]/H$\beta$ ratio compared to star formation, as opposed to Clumpy disks where the [O{\sc iii}]/H$\beta$ ratio can be AGN-dominated, leading to high Seyfert~2-like integrated probabilities.

\begin{table*} 
\centering
\begin{minipage}{\textwidth}{}
\centering
\caption{Optical line, X-ray and infrared luminosities for various sub-samples. We indicate mean values or values from mean stacking. Optical lines luminosities are uncorrected for dust extinction. Luminosities are in erg\,s$^{-1}$, except $L_{\rm IR}$ in $L_{\odot}$.
\label{tab:lum}}
\begin{tabular}{lcccc}
\hline
\hline
Mean luminosities
    &  Stable SF disks\footnote{\,Stable disk sample with $(1-P_{\rm MEx}(AGN))$ weights applied to each object to compute average values and to stack X-ray and infrared data, hence statistically selecting objects that have the highest probabilities to be purely star-forming.} 
    & All Stable disks\footnote{\,Entire Stable disk sample with equal weights for each object -- including the few with high AGN probabilities.}  
    & All Clumpy disks\footnote{\,Entire Clumpy disk sample with equal weights for each object -- including the few with low AGN probabilities.} 
    & Clumpy AGN disks\footnote{\,Clumpy disk sample with $P_{\rm MEx}(AGN)$ weights applied to each object to compute average values and to stack X-ray and infrared data, hence statistically selecting objects that have the highest probabilities to be clumpy AGN hosts.}  \\
\hline
$L_{\rm [O{\small III}]}$ & $2.1\pm0.5\times 10^{40}$          &  $2.9\pm0.4\times 10^{40}$ & $1.5\pm0.1\times 10^{41}$  & $2.1\pm0.2\times 10^{41}$   \\
$L_{\rm [H\beta]}$ & $3.3\pm0.4\times 10^{40}$                  & $3.5\pm0.4\times 10^{40}$    & $4.3\pm0.6\times 10^{40}$&  $4.1\pm0.5\times 10^{40}$ \\
$L_{\rm [O{\small II}]}$ & $3.2\pm0.8\times 10^{40}$                 & $3.8\pm1.1\times 10^{40}$    & $6.9\pm1.2\times 10^{40}$ &  $7.3\pm1.1\times 10^{40}$ \\
$L_{\rm X,\,0.5-8keV\,rest.}$ & $1.0\pm0.3\times 10^{41}$  & $1.2\pm0.3\times 10^{41}$  & $4.2\pm0.4\times 10^{41}$ &   $5.1\pm0.6\times 10^{41}$\\
$L_{\rm X,\,2-10keV\,rest.}$ & $0.6\pm0.4\times 10^{41}$  & $0.7\pm0.4\times 10^{41}$  & $2.6\pm0.8\times 10^{41}$ &   $3.0\pm0.8\times 10^{41}$\\
$L_{\rm IR}$ &         $5.0\pm0.5\times 10^{10}$ & $5.2\pm0.4\times 10^{10}$ & $4.5\pm0.6\times 10^{10}$ & $4.8\pm0.7\times 10^{10}$ \\
\hline
\end{tabular}
\end{minipage}
\end{table*}

{\revi
\subsection{Summary: optical line properties of Clumpy AGN candidates}

Clumpy galaxies show a much higher [O{\sc iii}] excitation than Stable disk galaxies in the same mass and redshift range. Comparisons with the properties of [O{\sc ii}] and [Ne{\sc iii}]  indicate that this trend is unlikely to be caused by metallicity effects in purely star-forming galaxies, and that the AGN fraction should be higher in Clumpy galaxies instead. Using the MEx diagnostic, not all Clumpy galaxies need to contain AGN, but many of them have a high AGN probability and the AGN probability correlated with both visual and quantitative estimates of the clumpiness. Similarly, some Stable disks may contain AGN, but the estimated fraction is significantly lower.

We list the average [O{\sc iii}], [O{\sc ii}] and H$\beta$ luminosities for both samples in Table~\ref{tab:lum}. Overall, Clumpy disks have an H$\beta$ luminosity higher by about 20\% than Stable disks, but their [O{\sc iii}] luminosity is higher by a factor five. This is not directly representative for the properties of the hosted AGN because some objects in the Clumpy sample are unlikely to host AGN, and vice versa. To better probe the properties of active clumpy disks, we have defined a ``Clumpy AGN'' sample, for which average properties are measured over our Clumpy disk sample with statistical weights equal to $P_{\rm Mex}(AGN)$, i.e. we give the highest weights to the systems that are the most likely to contain AGN. Conversely, we defined a ``Stable SF disk'' sample, which corresponds to the Stable disk sample with statistical weights proportional to the MEx probability of being purely star-forming. This allows a more relevant estimate of the properties of AGN hosted by Clumpy disks (see Table~\ref{tab:lum}). In particular, the typical [O{\sc iii}] luminosity of Clumpy AGNs is $2.1 \pm 0.2 \times 10^{41}$\,erg\,s$^{-1}$. This is ten times larger than in Stable SF disks, a result consistent with the more general properties of AGN hosts and star-forming galaxies in the SDSS \citep{kauffmann}, and suggesting that $\sim$\,90\% of the [O{\sc iii}] luminosity in Clumpy AGN comes from the AGN itself. 

Hence, the average AGN luminosity in the clumpy AGN candidates is $L_{[O{\sc iii}], {\rm AGN}} \approx 1.9 \pm 0.2 \times 10^{41}$~erg~s$^{-1}$ -- corrected for the minor contamination by star formation, but not corrected for dust extinction. This suggests an average bolometric AGN luminosity $L_{\rm bol, AGN} \approx 2 \times 10^{43}$~erg~s$^{-1}$ \citep[e.g.][]{netzer09} for the Clumpy disks with high AGN probabilities. 

The average H$\beta$ luminosity is slightly higher in the Clumpy AGN sample compared to the Stable SF one, by about 25\%. Given that Clumpy disks have slightly lower SFRs (Section~2.4), this indicates that about one third of the H$\beta$ luminosity in Clumpy AGN systems is powered by the AGN. The powering of the H$\beta$ luminosity by star formation is thus about similar in Clumpy and Stable disks, with an extra contribution to the H$\beta$ flux in Clumpy AGNs.
}


\section{X-ray stacking}\label{sec:stack}

\subsection{X-ray properties of individual sources and stacking}

AGN candidates in unstable disk galaxies are expected to have modest X-ray luminosities and significant obscuration by the high ISM column densities. They could thus be generally to faint to be individually detected even in deep {\em Chadra} surveys, especially in the hard band (B11). Indeed, only one object in our sample, C10, is a known hard X-ray selected AGN \citep{alexander03, xue11}. Several other sources are detected in the soft X-ray band of Chandra data (0.5-2keV), but in general with a signal-to-noise ratio that is too low to firmly distinguish between emission from an AGN or SF, both the soft X-ray luminosity and the star formation rate being relatively uncertain in these individual objects -- object C1 has a robust soft X-ray excess compared to its SFR, according to the Ranalli et al. (2003) relation, though.

\begin{figure*}
\centering
\includegraphics[angle=90,width=15cm]{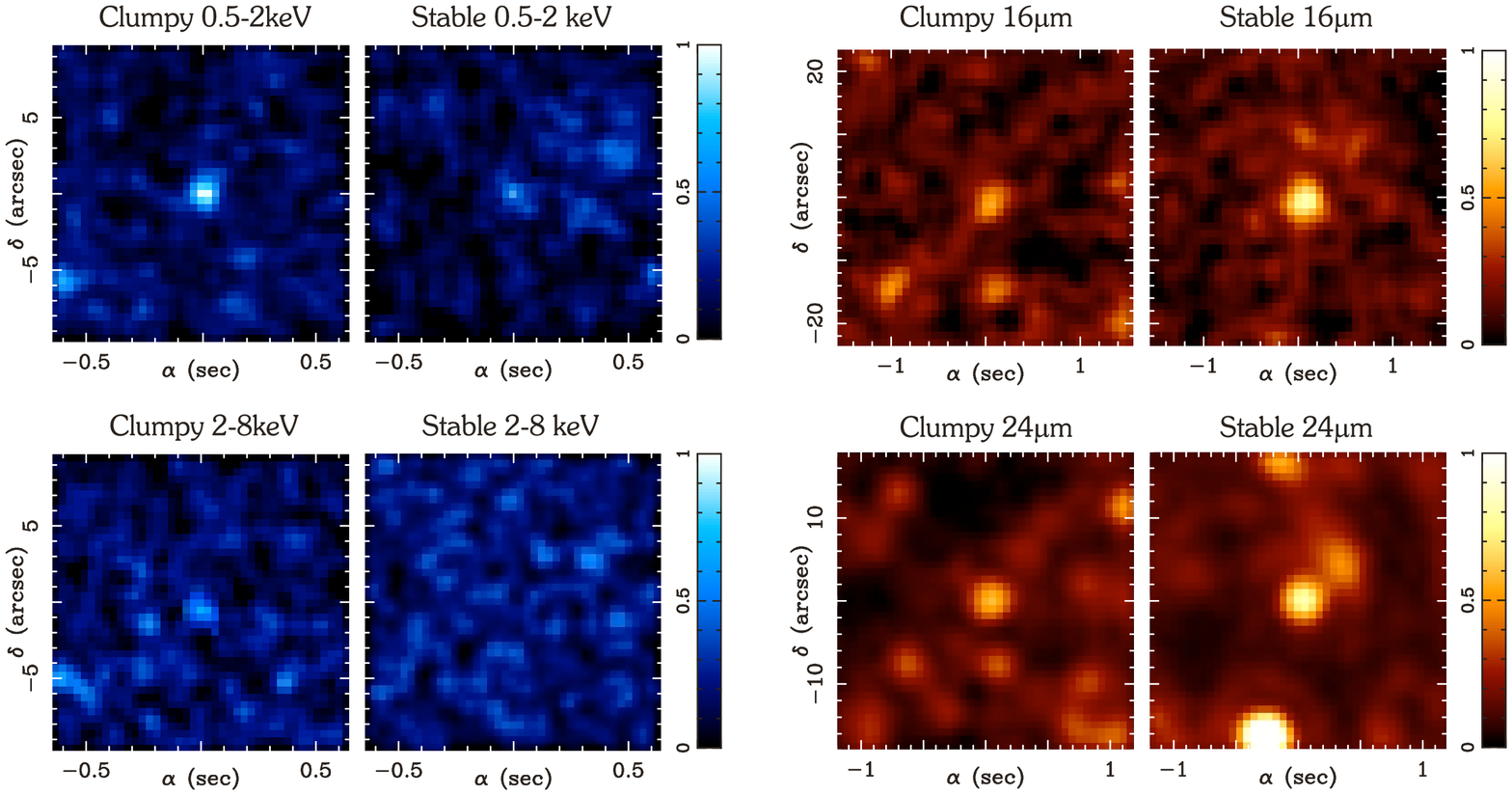}
\caption{Stacked data for the Clumpy and Stable disk samples, for soft X-rays, hard X-rays, 24$\mu$~m and 16$\mu$~m data, respectively. The stacked data were re-normalized to the same number of objects per sample, to the same luminosity distance for each individual object (see text for details), and background-subtracted. The same linear color bar is used for each pair of image, normalized to 1.0 for the brightest central pixel value for soft X-rays, 16 and 24~$\mu$m, and to 0.6 for hard X-rays. Note that the two bright areas at $\alpha \simeq 0.4-0.5$ sec in the soft X-ray stack for Stable disks are detections of nearby sources and not indicative of the typical noise level. The same applies in a more evident way to the 16 and 24~$\mu$m stacks with some nearby sources detected.
 \label{fig:xir}}
\end{figure*}

To confirm the high AGN frequency in clumpy galaxies, we performed mean X-ray stacking as in, e.g., \citet{worsley05} and \citet{daddi07}. All of our targets lie within 8' from the {\em Chandra } aim point in the CDFS, 80\% of the sources being within 6.5', and the median distance from the aim point being 4.3', where the sensitivity is relatively homogeneous (relative variations below 25\%). No target lies within the Chandra PSF of a close (in projection) X-ray source. For each source, the X-ray data were multiplied by a factor proportional to the square of the luminosity distance -- stacked results without this correction factor were also examined and yield similar results. The stacked data, presented in Figure~\ref{fig:xir}, were also normalized to the number of sources per sample.

\subsection{Stacked X-ray detections}

In the soft X-ray band (0.5--2~keV), X-ray stacking yields a 4.5-$\sigma$ detection\footnote{$\sigma$ being measured from the background pixel-per-pixel fluctuations} for the Stable disks sample, and a $3.7\pm0.2$ times larger signal for the sample of Clumpy galaxies. The stacked signal is not dominated by the few most luminous systems in soft X-rays: the results are similar if we select randomly only half of the galaxies in each sample. Hence the factor of $\approx \! 4$ soft X-ray excess is a general property of our whole Clumpy galaxy sample, not just of a few outliers within this sample. In the hard band (2--8~keV), there is no detection for Stable disks, and only a marginal detection for Clumpy disks. The stacked data show a 3.5-$\sigma$ detection but a single object (C10, which is the individually detected source) contributes 35\% of stacked the signal and rest of the sample gives only a 2.4-$\sigma$ signal.

There is thus a higher X-ray luminosity in Clumpy galaxies than in Stable disks. The hard/soft count ratio for Clumpy disks is $H/S \sim 6.5$ if the stacked signal in the hard band is robust -- conservatively, this is rather an upper limit. In the following we assume a spectral slope $\Gamma = 1.2 \pm 0.5$ for luminosity conversions, consistent with this (uncertain) $H/S$ count ratio and typical for AGN that are relatively obscured but not Compton Thick (e.g., Alexander et al. 2003, J11). The comparison of \oiii and X-ray luminosities in our sample is indeed consistent with such substantial obscuration (see Section~5.4). The stacking results indicate an average X-ray luminosity $L_{X (2-10 keV rest.)} = 2.6 \pm 0.8 \times 10^{41}$\,\egs~for Clumpy galaxies, versus $0.7\pm0.4 \times 10^{41}$\egs for Stable disks, both values being uncorrected for absoprtion (see Tabel~\ref{tab:lum}).


\subsection{Star formation rate and X-ray excess}

The X-ray luminosity of Clumpy galaxies is almost four times higher than that of Stable disks, but their star formation rates are not higher. They have somewhat higher sSFRs, but lower stellar masses too, and their absolute SFR was found to be slightly lower than that of Stable disks (from SED fitting and 24\,$\mu$m fluxes, Section~3.4, also consistent with H$\beta$ luminosities once corrected for AGN contribution, Section~3.7). 

SFR estimates from individual detections at 16 and/or 24\,$\mu$m are on average 7.5\sy for Stable disks, corresponding to $L_X/SFR\simeq 9 \times 10^{39}$\egs/\sy, suggesting X-rays in these systems trace mostly star formation \citep{ranalli03,grimm03}. In detail, this $L_X/SFR$ ratio is somewhat above the main trend defined by Ranalli et al. 2003, suggesting that there might be weak AGN in some of the Stable disks, consistent with the MEx results. The average SFR estimate in Clumpy disks, at 6.1\sy, results in $L_X/SFR \simeq 4 \times 10^{40}$\egs/\sy, indicating an X-ray excess in these systems, by a factor $> \,5$ compared to the X-ray luminosity expected from their star formation activity using the Ranalli et al. relation.

However, not all targets are individually detected at 16 or 24\,$\mu$m. Therefore we stacked the 16 and 24\,$\mu$m Spitzer data for our two samples of galaxies. Here we applied the same weighting with luminosity distance as for the X-ray stacking, but also checking that un-weighted stacking gives similar results, and normalizing stacked data to the same number of object per sample. The results are shown on Figure~\ref{fig:xir}: note that some infrared sources relatively close to our targets do appear on the stacked images, especially at 24\,$\mu$m, but none of our targets lies within the Spitzer PSF of another identified infrared source, and flux ratios between our two samples were measured within the FWHM of the PSF to avoid contamination.

The results clearly show that there is no 24 or 16\,$\mu$m excess in Clumpy galaxies. Instead, their stacked signal is $32\pm4$\% lower at 16\,$\mu$m and $23\pm3$\% lower at 24\,$\mu$m than that of Stable disk galaxies. 24\,$\mu$m are not proportional to star formation rates in a redshift-independent way. Here the main bias between the two samples would result from the highest-redshift objects being Clumpy ones at $z \simeq 0.85$, where PAH lines are redshifted in the Spitzer 24\,$\mu$m band and increase the observed flux at fixed SFR. This may explain why the Clumpy-to-Stable flux ratio is somewhat higher (smaller than unity but closer to unity) at 24\,$\mu$m than at 16\,$\mu$m -- a small effect that could only strengthen our conclusion.

\smallskip

{\revi Thus, star formation cannot explain the X-ray excess in Clumpy disk galaxies. One could still envision that Clumpy galaxies have an X-ray excess for some reason else than hosting AGN, for instance shocks in the ISM\footnote{There is probably strong supersonic turbulence and shocks in clumpy disks, however, spiral arms in Stable disks typically correspond to strong shocks in interstellar gas, too.}. However, \citet{ranalli12} found that the local X-ray relations for star-forming galaxies hold at increasing redshift, where clumpy types should be increasingly frequent. 

To further test the consistency of the AGN interpretation of the X-ray excess with the previous optical line results, we performed the X-ray stacking measurements for the Clumpy galaxy sample with statistical weights proportional to $P_{\rm Mex}(AGN)$ (previously defined as the "Clumpy AGN" weighted sample) and the corresponding "Stable SF" disk sample with weights proportional to $(1-P_{\rm Mex}(AGN))$. As indicated in Table~\ref{tab:lum}, the average X-ray luminosity increases in the weighted Clumpy AGN sample, compared to the un-weighted Clumpy disk sample, while SFR estimates ($L_{\rm IR}$, $L_{H\beta}$, SED fitting) remain relatively unchanged. Hence, the X-ray excess in Clumpy disks is strongest in objects that have the highest probability to host AGN based on their nebular line properties, as expected if the X-ray excess is indeed caused by frequent AGN in Clumpy disks. This further demonstrates the agreement of the two methods (X-ray stacking and optical lines) toward the AGN interpretation.
}

\subsection{Inferred AGN and BH properties}

The identification of AGNs in Clumpy galaxies with various techniques indicates that the AGN dominates the [O{\sc iii}] line emission and X-ray emission, on average, in this galaxy sample. Here, we discuss the typical BH mass accretion rate from the observed AGN luminosity although the interpretation of \oiii  and X-ray luminosities is subject to large uncertainties given the potential effects of obscuration.

\paragraph{Luminosity and obscuration}

The average X-ray luminosity of the Clumpy AGN sample (weighted using the MEx probabilities to be as representative of actual AGN hosts as possible) is $L_{\rm X,\,2-10keV\,rest.} \approx 3 \times 10^{41}$\egs, after estimating and subtracting the contribution of star formation with the relations from \citet{ranalli03}. Direct conversion into AGN bolometric luminosity from the X-ray data alone is uncertain as the obscuration cannot be measured using a hard band to soft band ratio for our sample.

The comparison with [O{\sc iii}] luminosities in Table~\ref{tab:lum} shows that $L_{X}/L_{\rm [OIII]} \leq 2$, which means that these AGN could be substantially obscured\footnote{because unobscured AGNs would have $L_{X}/L_{\rm [OIII]} \geq 10$} but not necessarily Compton thick. This indicates that the intrinsic X-ray luminosity of these AGN would be higher than observed, and we infer from the observed \oiii  luminosities an intrinsic X-ray luminosity in the range of $L_{X,{\rm int.}} \sim 10^{42-43}$\egs \citep{netzer06}. This is consistent with the models in B11, predicting high gas column densities ($N_{H} \sim 10^{23-24}$~cm$^{-2}$ from the general large-scale ISM, not counting extra obscuration from a potential AGN torus) on most lines of sight but reaching Compton thickness on a fraction of these. This is also consistent with the observation by Mullaney et al. (2011) of numerous X-ray AGN on the Main Sequence of star formation that could be obscured without being Compton thick.

\paragraph{Bolometric luminosity and BH accretion rate}

Estimating the AGN bolometric luminosity from the observed [O{\sc iii}] luminosity is quite uncertain. For instance, the conversion factor depends strongly on the X-ray luminosity corrected for absorption (e.g., Netzer et al. 2006), which is uncertain here, and the $L_{\rm bol}/L_{\rm [OIII]}$ values reported in the literature span a range from a few tens to several thousands. If we adopt a relatively conservative ratio for MEx-selected AGN of $\approx$300, the average AGN bolometric luminosity is of the order of a few $10^{43}$\egs. Some studies have suggested higher conversion factors for similar [O{\sc iii}] luminosities, up to $\approx$3000 \citep{netzer09,heckman04, H05, shen}, in which case the AGN bolometric luminosity may reach a few $10^{44}$\egs.

If the mass to luminosity conversion efficiency is 10\%, the AGN luminosity is $L_{\rm bol} \sim 10\%\, \dot{m}\, c^2$, then the average BH accretion rate is $\dot{m} \sim 10^{-2} - 10^{-1}$\sy, depending on the adopted conversion factors. For a star formation rate of a few \sy, this corresponds to an $\dot{m}/SFR$ ratio around $10^{-3}$ or somewhat larger, as found more generally for BH growth in gas-rich galaxies at $z>1$ \citep{daddi07, mullaney12} and implying that these BHs could be growing toward the observed scaling relations \citep{magorrian,MF01}. If the violent disk instability process can be maintained over timescales of a few Gyr (or even less for the highest estimates of the bolometric luminosity), the resulting BH mass could be a few $10^{7}$\ms, possibly representing most of the final BH mass in galaxies with stellar masses of a few $10^{10}$\ms. The average AGN luminosity remains sub-Eddington (even for the highest [O{\sc iii}]-to-bolometric luminosity conversion factors), but short accretion bursts with higher Eddington ratios could be possible, for instance if giant clumps migrate and coalesce centrally.


\section{Discussion: comparison to higher-redshift data and models}

Our sample of clumpy galaxies at $z \! \sim \! 0.7$ has properties that are globally representative of gas-rich unstable disks, which are increasingly common at higher redshift. As detailed in Section~2, their relatively high but Main Sequence star formation rate are indicative of disks with high gas fractions rather than mergers. This suggests that the results obtained for our intermediate-redshift sample are also representative of the numerous unstable clumpy galaxies at $z \! > \! 1$, most of which are gas-rich systems on the Main Sequence. At redshift $z \! \approx \! 2$, \citet{guo11} observed in the Hubble Ultra Deep Field that some clumpy disk galaxies contain X-ray detected AGN. Having only a limited fraction of such AGN individually detectable in X-rays is fully consistent with our findings. Our narrow line and X-ray stacking results actually indicate that there should be more AGN of this type than those unveiled by individual X-ray detections. In massive disks at $z \! \approx \! 3$, \citet{cresci10} noted that the [O{\sc iii}]/H$\beta$ ratio sometimes peaks in inner regions of the disk, either at the center of close to it, not in the outer disk as expected for the usual inward metallicity gradients. In the light of our results, this might be another signature of AGN activity in clumpy rotating disks.

The triggering of nuclear activity and BH growth by violent disk instabilities should then be a general process in stream-fed, gas-rich, unstable disk galaxies at any (higher) redshift, and in particular at the peak epoch of cosmic star formation activity around redshift two. Our estimates of AGN luminosities and BH accretion rates are consistent with the theoretical expectations from B11, and over timescales of one to a few Gyrs this process could fuel a large part of the final BH mass in moderate-mass galaxies, and potentially even in massive objects. Hence, the violent clump instability in high-redshift galaxies is much more efficient in feeding a BH than the weaker bar instabilities frequent in low-redshift systems. However, bars are the most considered instability mode in theoretical models to date \citep[e.g., ][]{fanidakis11}. Another difference is that high-redshift clump instabilities can grow a BH together with either a classical bulge or a pseudo-bulge \citep{noguchi,EBE08b,saitoh}, while bars could only grow pseudo-bulges.

As disk galaxies fed by rapid cold gas accretion could remain unstable for a few Gyr, the duty cycle of BH growth could be high ($>$10\%) in this phase. This does not require BHs to be permanently active in these galaxies: some Clumpy disks in our sample have low AGN probabilities and/or no X-ray excess with respect to their SFR. As an illustration, if we assume that $\sim$50\% of Main Sequence star-forming galaxies at $z$=1-2 are clumpy disks (consistent with deep imaging and spectroscopic surveys, e.g. Elmegreen et al. 2007 and Shapiro et al. 2008), and 50\% of them have AGN with $L_{X} \geq 10^{42}$\,\egs, this could contribute to an AGN fraction of roughly one fourth in Main Sequence star-forming galaxies (considering AGNs with an intrinsic $L_{X}$$\geq$$10^{42}$\,\egs). Our data and the models in B11 suggest that clumpy disks can host more luminous AGNs, maybe even at Eddington-limited levels, but with lower AGN fraction and duty cycles for such higher-luminosity populations.

\medskip

These findings could explain how AGN are frequently fed in galaxies that do not display the signatures expected for major mergers \citep{grogin05,gabor09,kocevski11,mullaney11}. In this long duty cycle mode, with ubiquitous instabilities in high-redshift disks fed by rapid gas accretion, a large part of the BH mass in today's galaxies could have been assembled in AGN phases that have moderate luminosities and significant obscuration, most often with bolometric luminosities of $\sim$\,$10^{43-44}$\,erg\,s$^{-1}$. Short and bright phases could occur when dense clumps reach the nucleus and the AGN luminosity reaches QSO levels and the BH accretion rate increases, potentially reaching Eddington-limited levels (B11). In this case, each given system would spend most of the time at moderate-luminosity AGN levels, but it would be possible that a larger fraction of the BH mass is assembled during shorter high-luminosity phases (or not). This is impossible to study in our small sample, but is suggested by some observations \citep[although lacking information on the nature of host galaxies -- e.g., ][]{soltan} and some cosmological simulations \citep[although not resolving the detailed accretion processes yet -- e.g., ][]{martizzi}.

The contribution of violent disk instabilities to the luminosity function of AGNs and mass function of BHs remains to be established. Here we have provided evidence that high-redshift disk instabilities trigger BH growth: this can be through BH formation in the giant clumps (Elmegreen et al. 2008), through giant clumps migration bringing gas directly into a central BH, or through the more general gas inflow that giant clumps trigger in high-redshift disks {\em even} when they don't reach the center themselves (Bournaud et al. 2011). This is similar to the known and observed triggering of BH growth by interactions and mergers (e.g., Ellison et al. 2011). But although mergers do trigger BH growth, selections of AGN host galaxies do not find a substantial excess of mergers with respect to control sample, implying that merger-induced process does not dominate the triggering of AGN in the Universe. The process of violent instability resulting from rapid cold gas accretion seems to dominate the stellar mass assembly of galaxies compared to big mergers (Dekel et al. 2009, Brooks et al. 2009, L'Huillier et al. 2011), so this process could be more likely to significantly contribute to AGN triggering. Kocevski et al. (in preparation) find that clump instabilities are not a dominant trigger of AGN with observed X-ray luminosities above $10^{42}$\,\egs. This remains consistent with our present findings, as our sample of fourteen Clumpy disks would include only one such AGN, with numerous AGN revealed by other means -- narrow line diagnostics, X-ray stacking, and in a few cases a possible X-ray excess compared to the SFR but below $10^{42}$\,\egs.


\section{Conclusion}

The growth of Supermassive Black Holes (BHs) triggered by disk instabilities in high-redshift galaxies could be hard to observe directly at redshifts $z \! > \! 1$, especially in X-rays.  According to numerical simulations, this mode would, most of the time, trigger moderate luminosity and obscured AGNs. In this paper, we have built a sample of unstable (Clumpy) disks and a control sample of Stable disks at intermediate redshift ($z \! \sim \!  0.7$). This way, more reliable Narrow Line diagnostics can be used compared to higher redshift, and moderate X-ray luminosities are in easier reach of (stacked) observations. Our classification of Clumpy and Stable disks used visual estimates of the clumpiness and was confirmed by automated measurements. The properties of the selected Clumpy galaxies are consistent with the expectations for gravitationally unstable disks rather than mergers, in particular specific star formation rates (sSFR) on the Main Sequence of star formation rather than in a starbursting mode\footnote{although a large fraction of mergers are also on the Main Sequence \citep{kartaltepe}.}. Velocity fields, available for some of these objects in the Literature, are consistent with rotating disks. Thus, we have compared a sample of violently unstable disk galaxies, similar to many $z \! > \! 1$ star-forming galaxies, to a sample of more stable, spiral-like disk galaxies, with lower gas fraction and weaker instabilities.

Utilizing the Mass-Excitation (MEx) diagnostic introduced by Juneau et al. (2011), {\em we find that violently unstable (Clumpy) systems frequently host an AGN, which is rarely the case in more stable galaxies}. Comparisons of results from the MEx and from other narrow-line diagnostics indicate that the much higher \oiiihb  ratios observed in clumpy systems cannot simply result from lower gas-phase metallicity, and actually require frequent Seyfert~2-type activity. X-ray signatures are in general too weak to be securely identified in individual systems -- although one system has a hard X-ray detection and a few cases have a soft X-ray excess w.r.t star formation. Stacking reveals a clear X-ray excess in clumpy disk galaxies, which significantly exceed the X-ray contribution from star formation. This confirms the identification of frequent AGN in these systems. Intrinsic AGN bolometric luminosities are conservatively estimated to be of a few $10^{43}$\,\egs and potentially of a few $10^{44}$\,\egs, in Clumpy disks with stellar masses of a few $10^{10}$\,\ms~at $z$\,$\sim$\,0.7. We also find that these AGN are, on average, substantially obscured in X-rays, but not necessarily Compton thick.

\medskip

Clumpy disk galaxies appear to have higher gas fractions than stable disks, but by a factor lower than two in our mass-matched and redshift-matched samples. Thus, Clumpy disks contain somewhat larger gas reservoirs\footnote{as expected to trigger their instability}, but not to the point of directly explaining the strong enhancement of the \oiii and X-ray luminosities by a factor of about five. An associated dynamical process for AGN feeding, triggered by the presence of clumps, is thus required. It may be either black hole growth in the giant clumps (Elmegreen et al. 2008) and/or gas inflows toward a central BH triggered by the presence of giant clumps -- which can sometimes consist in clump migration and central coalescence, but should be triggered even when the clumps are not merging centrally (B11). We cannot specify the location of the AGN in our sample as we did not use spatially-resolved tracers. The average AGN luminosity is about five times higher in our Clumpy disk sample than in the Stable disk control sample, although this cannot directly constrain the AGN fraction among clumpy galaxies in general, since our sample was emission line-selected. Given the frequency of Clumpy disk instabilities at high redshift, and the typical AGN luminosity of at least few $10^{43}$\,\egs~ in many of these systems, a large part of today's BH mass in galaxies with stellar masses of $10^{10-11}$\,\ms~could be fueled in this process.

\medskip

Violent disk instabilities should persist longer in lower-mass galaxies before being replaced by weaker secular modes (bars and spirals). This has been tentatively observed by \citet{E09}, is supported by our own sample (see Section~3.4), and is also found in samples of zoom-in cosmological simulations \citep{martig12, kraljic12}. This is naturally explained by the evolution of cold accretion with redshift and mass \citep{DB06} and of gas consumption with mass and metallicity \citep{KD11a}, both of which eventually result in disk stabilization \cite{cacciato}. If violent disk instabilities in galaxies fed by cold gas accretion drive the bulk of the cosmic SF history, a downsizing in the termination of star formation should result. i.e. lower-mass galaxies continue to form stars actively at lower redshifts, as observed \citep[e.g.,][]{J05}. The downsizing of BH growth \citep[e.g.,][]{hasinger,labita} could also naturally result from the fueling by violent disk instabilities at high redshift in high-mass galaxies, and down to intermediate-redshift in lower-mass galaxies, as a natural consequence of the mass and redshift dependence of disk instabilities.



\appendix

We describe in the appendices the detailed methods used for stellar mass and emission line ratio measurements.

\section{Stellar masses}\label{sec:mass}

The stellar masses used in this work were derived from $K$-band luminosities \citep{isaac-goods}, assuming a redshift-dependent mass-to-luminosity ratio for star-forming galaxies given by \citet[][equation~2]{arnouts07}. A correction factor of 0.2~dex was applied to convert stellar masses from a Salpeter IMF (as used in \citealt{arnouts07}) to a \citet{chabrier} IMF (as appropriate for the MEx diagnostic, J11).

We performed SED fitting for objects with reliable luminosities from the near-ultraviolet to mid-infrared in the FIREWORKS catalogue \citep{wuyts}. This was done using the \citet{BC03} stellar population models, and assuming exponentially-decaying single-burst star formation histories (so-called tau-models). We used stellar ages from 10~Myr to 13.5~Gyr, and a Calzetti extinction law with E(B-V) from 0.0 to 0.7. The SED fitting results yielded stellar masses lower by 0.2-0.3~dex than those obtained using a redshift-dependent $M/L_K$ ratio (for both smooth and clumpy disks). This is in fact consistent with the findings by \citet{maraston10} on the effects of tau-models on stellar mass estimates for star-forming galaxies: using such star-formation histories is expected to under-estimate the real stellar mass by 0.2-0.3~dex. For this reason, we preferred to use a redshift-dependent $M/L_K$ ratio. Indeed, \citet{bitsakis} have shown, using the \citet{dacunha} SED models, that redshift-dependent $M/L$ ratios is in good agreement with the results from SED fitting when the assumed star formation histories are not limited to tau-models.

We note that a more recent calibration of redshift-dependent $M/L_K$ ratios was proposed more recently by \citet{ilbert10}, but based only on tau-models and hence likely to under-estimate the stellar mass of star-forming galaxies . We thus used the calibration from \citet{arnouts07}, which is based on composite star-formation histories and, in the $0.5<z<1.0$ range, is calibrated on galaxies in a stellar mass range very close to that of our samples. 

\medskip

We attribute a uniform uncertainty of 0.3~dex to all stellar mass estimates, because of the uncertainties on the $M/L_K$ calibration (0.21~dex) and $K$-band luminosities ($<$0.15~dex). Note that even SED fitting with tau-models would yield results within this uncertainty. Stellar masses were estimated independently for objects C7 and S5 by \citet{rodrigues08}, and the independent estimated from these authors are very similar to ours (within 0.1~dex, i.e., well within the error bar).

\medskip

We also checked consistency with dynamical masses in two cases, using velocity fields published in \citet{yang08}. Object S3 has a rotation velocity $V \simeq 160$\kms for an optical radius $R\simeq 14$~kpc and optical inclination $i \sim 45^\circ$, indicating the dynamical mass $M_\mathrm{dyn} = R\,V^2/G$ within the optical radius. Assuming that two thirds of the mass within this radius is baryonic \citep[e.g.,][]{FS06} and that one third of the baryons in $z$$\sim$0.7 star-forming galaxies are in the form of gas \citep{tacconi10}, this gives a stellar mass $\log(M*_/M_\odot)=10.8$. For object, C7, $R \simeq 8$~kpc, $V\simeq 105$\kms, $i \simeq 50^\circ$, which results in $\log(M*_/M_\odot)=10.1$. These dynamical estimates ($\log(M*)$=10.8 and 10.1 for S3 and C7, respectively) are consistent with our measurements ($\log(M*)$=10.9 and 10.2, respectively) within the assumed uncertainty of 0.3~dex.

\section{Emission line ratios}\label{sec:line}

We checked in the ESO database that the slit used during the observations includes at least 75\% of the B-band emission for each selected system -- in general, the whole galaxy is covered. We measured emission line fluxes by fitting a Gaussian function on lines. 

A linear continuum was estimated on a 80\AA range on each side of the line, which provides a large enough range after excluding sky lines, other emission lines, and instrumental features. The flux was measured by integrating the fitted Gaussian on the wavelength range over which it exceeds $2.5 \sigma$, where $\sigma$ is the noise level estimated on the continuum. Given the measured noise level, we assumed a uniform uncertainty of 15\% on measured fluxes and 0.2~dex on line ratios. Alternative flux measurement achieved by integrating lines over the range where the exceed $2.5 \sigma$ without using a Gaussian fit give results that lie within this uncertainty.

Underlying \hb absorption could be robustly fitted in almost half of the cases. The resulting correction to the \hb flux was a factor ranging from 1.09 to 1.21, with a median value of 1.13. We applied a constant 1.20 to all spectra, which is conservative as it tends to (slightly) lower the AGN probabilities on the MEx diagnostic for Clumpy disks, as the average \hb absorption is in fact stronger for Stable disks than for Clumpy disks (see Section~\ref{stack}). H$\delta$ absorption can be measured on stacked spectra for clumpy and stable disks samples independently (see Section~\ref{stack}): it implies an \hb absorption correction factor of 1.16 for stable disks and 1.12 for clumpy disks, so we tend to slightly over-correct the \hb flux for clumpy disks, but in a way that does not alter the results and anyway remains within the assumed uncertainty. These potential differences in \hb correction between Clumpy and Stable disks, compared to the 1.20 correction factor applied to all spectra in our analysis, correspond to an offset smaller than 0.05 dex on the MEx diagram, not changing significantly the estimated AGN probabilities.
 
We note that \oiiihb and \oii/\hb ratios were measured by \citet{rodrigues08} from a different spectroscopic survey for three of our targets, and are consistent with our own measurements and the assumed uncertainties.

\acknowledgments

We acknowledge useful and insightful discussions with David Alexander, Damien Le~Borgne, Elisabete Da~Cunha, Bruce Elmegreen, Debra Elmegreen, Eric Emsellem, Natascha F\"orster-Schreiber, Dale Kocevski, Matthew Lehnert, Mark Sarzi and Romain Teyssier, and useful comments from an anonymous referee. We acknowledge support from the EC through grants ERC-StG-257720 (FB, SJ), the CosmoComp ITN (FB, SJ), grants ISF 6/08, GIF G-1052-104.7/2009, NSF AST-1010033 and a DIP grant (AD).

\end{document}